\documentclass[aps,prb,twocolumn,amsmath,amssymb,superscriptaddress]{revtex4}
\usepackage{graphicx}
\newcommand{\ud}{\mathrm{d}}
\newcommand{\vc}[1]{\mathbf{#1}}
\newcommand{\real}{\mathrm{Re}}

\newcommand{\infinity}{\infty}

\begin{document}


\title{Quantum Vortex Dynamics: Results for a 2-d Superfluid }


\author{Timothy Cox}
\affiliation{Department of Physics \& Astronomy, University of British Columbia, Vancouver, BC V6T 1Z1, Canada}
\author{Philip C. E. Stamp}
\affiliation{Department of Physics \& Astronomy, University of British Columbia, Vancouver, BC V6T 1Z1, Canada}
\affiliation{Pacific Institute of Theoretical Physics, University of British Columbia, Vancouver, BC V6T 1Z1, Canada}


\date{\today}

\begin{abstract}

We model vortex dynamics in a 2-dimensional Bose superfluid using the Thompson-Stamp (TS) equations of motion, which describes both the classical Hall-Vinen-Iordanskii (HVI) dynamical regime and the fully developed quantum regime, and the crossover between them. The TS equations can be written in the form of a quantum Langevin equation. Analytic solutions are given for all of these regimes for a single vortex in a 2-dimensional system. In the classical regime we include the vortex inertial and Langevin noise terms, which are dropped in the usual HVI analysis. In the quantum and crossover regimes the effect of memory terms is important, and leads to clear differences from the classical predictions.

\end{abstract}


\maketitle

\section{Introduction}
One of the most fundamental problems in the theory of superfluids is to find the correct description of quantum vortices and their dynamics. In $^4$He superfluid, the first superfluid to be discovered and investigated, the debate over the correct way to describe vortex dynamics has now been going on for 60 years. More recent applications have been to both superconductors and Bose gas superfluids. Quite apart from any applications, there are also key theoretical principles involved, which demand a correct theoretical description of vortices in any superfluid - this has led to extended controversy over the years. 

In the original work of Hall and Vinen \cite{HallVinen} and Iordanskii \cite{Iordanskii64}, a phenomenological equations of motion, the HVI (Hall-Vinen-Iordanskii) equation, was used to describe the dynamics of a single superfluid vortex. These equations are a strictly local and classical equations for the vortex coordinate ${\bf R}_v(t)$. Once an inertial mass term, absent from the original papers \cite{HallVinen,Iordanskii64}, is added, it is most simply written for a vortex moving in a 2-dimensional superfluid, and there takes the form:
\begin{align}
 \label{eom1}
M_v \ddot {\bf R}_v - {\bf f}_M - {\bf f}_{qp}^{(cl)} - {\bf F}_{ac}(t)
\;=\; 0
\end{align}
where ${\bf F}_{ac}(t)$ is an applied driving force, and $M_v$ is the vortex hydrodynamic inertial mass. There is also a Magnus force ${\bf f}_M$ and a force ${\bf f}_{qp}^{(cl)}$ generated by the normal quasiparticle fluid. The HVI equations are strictly local in space and time - the forces act locally on the vortex, and there are no time retardation effects.

The HVI equations of motion have been applied with considerable success to a variety of superfluid and superconducting systems (for which see several reviews \cite{Sonin87,Blatter94,Sonin97,Kopnin02}). Nevertheless, as noted above, strong debate has attended its use. The modern discussion began with the work of Thouless et al. in the early 1990's \cite{ThoulessI}, and has continued since then \cite{ThoulessII}. In the work of Thouless and collaborators, a strictly local equations of motion were still advocated, but the nature and magnitude of some of the forces in the HVI equations was questioned. Much of the discussion in the period from 1990-2007 centred on the existence or otherwise of the Iordanskii force, and its magnitude and temperature dependence; the definition and size of the vortex effective mass was also discussed in many papers \cite{Popov73,Duan94,Arovas97,ThoulessII,MacDonald10}. Critical analyses of the Thouless et al. work \cite{Volovik,Sonin98,Wexler98} debated the validity and use of the HVI equations, but no clear resolution was reached.

However there are 3 important problems that were not touched upon in this debate, and which were then raised by Thompson and Stamp \cite{TS1,TS2}. These are:

(i) The HVI equations are entirely classical - yet the vortex is a quantum object, and should properly be described by quantum-mechanical equations of motion;

(ii) the HVI equations are local in space and time. This is physically unreasonable, given that the vortex is interacting with quasiparticle excitations, which can be emitted or scattered and then interact later with the vortex, and which also cause non-local backflow effects - all of this should then lead to non-local 'memory' effects, in both space and time, in the vortex equations of motion; and

(iii) Even if a local description were accurate, one would still expect to see a Langevin-type noise term in any vortex equations of motion, in addition to the forces in (\ref{eom1}).

In an effort to resolve these issues, as well as those raised by Thouless et al., the work of Thompson and Stamp \cite{TS1,TS2} derived the full quantum equations of motion for the vortex reduced density matrix $\bar{\rho}({\bf R},{\bf R'};t)$, obtained by integrating out all the quasiparticle degrees of freedom; here the vortex coordinate states
$|{\bf R} \rangle, |{\bf R'} \rangle$ are defined by the position of the vortex node appearing in the many-body wavefunction.  Then, by defining a vortex ``centre of mass'' coordinate ${\bf R}_v = {1\over 2}({\bf R}+{\bf R}')$, and a ``quantum fluctuation coordinate'' $\boldsymbol{\xi} = {\bf R}-{\bf R}'$, one can extract a pair of quantum Langevin equations, for both ${\bf R}_v(t)$ and $\boldsymbol{\xi}$.

For the centre of mass coordinate ${\bf R}_v(t)$ one can write the TS equations of motion in the form of a force balance equations, as follows:
\begin{align}
 \label{eom2}
M_v \ddot {\bf R}_v - {\bf f}_M - {\bf f}_{qp} - {\bf F}_{ac}(t)
\;=\; {\bf f}_{fl}
\end{align}
In this equation the inertial, Magnus, and applied force terms are unaltered; however the quasiparticle force ${\bf f}_{qp}$ now becomes non-local in both space and time, having components
\begin{equation}
f_{qp}^{\alpha} \;=\; \int_{-\infty}^t ds \gamma^{\alpha \beta} (t-s) \dot{R}_v^{\beta}(s)
 \label{f-qp}
\end{equation}
(with terms depending on both $(\dot{\bf R}_v(s) - {\bf v}_n({\bf R}_v))$ and $(\dot{\bf R}_v(s) - {\bf v}_s({\bf R}_v))$, where ${\bf v}_s, {\bf v}_n$ are the velocities of the superfluid and normal components of the system); and we also have a fluctuating (quantum Langevin) force ${\bf f}_{fl}(t)$ driving the dynamics. The general form of these forces is given below; the key features we wish to emphasize here (and which we elaborate on below) are:

(i) the memory effects in the non-local quasiparticle force ${\bf f}_{qp}(t)$ are small in the classical regime (this regime will be defined below) but become very pronounced in the quantum regime, to the extent that they severely modify the vortex dynamics; and

(ii) the fluctuation force ${\bf f}_{fl}(t)$ has a simple Markovian behaviour in the classical regime, but is strongly non-Markovian in the quantum regime (ie., it is also strongly non-local in time).

There is also an equation of motion for the quantum fluctuation coordinate $\boldsymbol{\xi}(t)$, also strongly non-local, which is of some interest for the general theory - however in this paper we will only be concerned with the dynamics of ${\bf R}_v(t)$.

The Thompson-Stamp (TS) equations are generally valid provided (a) the vortex velocity $\ll c_o$, the sound velocity, and (b) the characteristic energies in the vortex dynamics do not exceed the effective UV cutoff energy $\hbar \Lambda_o$ of the theory, which for a Bose liquid is given by $\hbar \Lambda_o =  m_o c_o^2$, where $m_o$ is the mass of the particles. There are 2 characteristic energies involved in the vortex dynamics - these are the thermal energy $\hbar \omega_T = k_BT$ of the quasiparticles, and $\hbar \Omega$, where $\Omega$ is the characteristic frequency of the vortex dynamics, itself typically controlled by the boundary conditions and/or the frequency of any applied force ${\bf F}_{ac}(t)$.

The upshot of these investigations can be summarized by saying that the HVI equations, and elaborations of them including inertia and noise, describe a {\it classical regime} of vortex dynamics, essentially a low frequency/high temperature regime with $\hbar \omega \ll k_BT$, where the the temperature $T$ refers to the normal quasiparticle fluid, and the frequency $\omega$ is that of the vortex dynamics (this being controlled both by the frequency of any driving force ${\bf F}_{ac}(t)$, and by the geometry of the sample). By lowering the temperature and/or increasing the vortex frequency, we cross over into a quantum regime with $\hbar \omega > k_BT$, where the vortex dynamics has very different properties, in which highly non-Markovian memory effects are crucial.

It is clear that it will be of considerable interest to see these results applied to real superfluid systems. However another important requirement for any application is to actually solve the equations of motion - this is what we will do in the present paper. Our main goals are (i) to solve the TS equations of motion, and compare the solutions to those of the HVI equations; and (ii) to give solutions in regimes that are accessible experimentally, so that they may be checked. Since our goal is to give analytic solutions, we will focus on the case of a single vortex in a 2-dimensional superfluid; the results in 3 dimensions are far too complex to be given in an analytic form. 

The plan of the paper is as follows: in section \ref{sec:EoM} we begin by reviewing the semiclassical equations of motion derived by Thompson and Stamp \cite{TS1}, and in section \ref{sec:Derivation} we derive the formal solution to the full equations of motion - in its full form it is a stochastic equation, because of the non-Markovian noise terms. In the rest of the paper we unravel this formal solution, looking in turn at the effects of noise, inertia and memory on the motion of the vortex. Thus in section \ref{sec:HVI} we discuss vortex motion in the HVI equations with noise; then we examine the effect of including inertia in section \ref{sec:HVIin}, and memory in section \ref{sec:HVImem}, and finally with both inertia and noise in section \ref{sec:HVImemin}.

In the present paper we will not try to give any application to experiments on particular systems - the idea is to first gain a detailed quantitative understanding of the vortex dynamics in the various regimes of interest.


\section{Vortex Equations of motion: General Features}\label{sec:EoM}


In what follows we give a detailed characterization of the various terms in the TS equations, comparing them where relevant to the analogous terms in the HVI equations. We conclude by rewriting the TS equations in a form which makes them amenable to analytic solution.

\subsection{Different terms in the equations of motion}
 \label{sec:Eom-item}

We start by dealing term by term with all the forces in the TS/Langevin equations, comparing where relevant with forces in the HVI equations.

\vspace{2mm}

{\bf (i) Vortex Inertial mass:} In all discussions of a vortex effective mass to date (and there have been a great many of these), a local inertial mass term has been employed. The value of this mass is typically derived using a hydrodynamic description - it then depends quite generally on the sample geometry and the position of the vortex in this geometry, ie., it has a non-local spatial dependence. For a vortex near the centre of a cylinder of radius $R_o$ the hydrodynamic vortex mass is well known to be \cite{Duan94,Popov73}:
\begin{eqnarray}
M_v & \;=\; &\pi\rho_sa_0^2\left[\log\left(\frac{R_o}{a_0}\right)+\gamma_E+\frac{1}{4}\right] \nonumber \\
& \; \equiv \; & \pi\rho_sa_0^2\log\left(\frac{R_r}{a_0}\right),
 \label{Meff}
\end{eqnarray}
where $\gamma_E$ is Euler's constant - this result is easily derived by considering the kinetic energy of the superfluid carried by the vortex. There are many treatments in the literature of the hydrodynamic mass for different sample geometries.

As noted elsewhere, results like this are not strictly correct - not only should the mass depend on the detailed motion of the vortex (in particular, it should be frequency-dependent), but there should also be non-inertial corrections to it of a similar nature to those arising in classical electrodynamics (involving 3rd and higher time derivatives in the equations of motion). The reason for the frequency dependence is intuitively obvious - a motion of the vortex over a timescale $\Delta t$ can only affect superflow within a radius $\Delta r \sim c_o \Delta t$ of the vortex, where $c_o$ is the sound velocity. Thus we expect the vortex mass to decrease logarithmically with frequency, for frequencies $\omega > \omega_o \sim c_o/R_o$, in a container of size $R_o$.

In this paper we will ignore these issues, since our main objective here is to understand the role of the non-inertial forces acting on the vortex. Nevertheless, all result derived below containing the inertial mass of the vortex should be treated with appropriate caution.

\vspace{2mm}

{\bf (ii) The Magnus force:}  The Magnus force in superfluids has also been widely discussed over the years. It takes the classic form
\begin{equation}
{\bf f}_M = \rho_s \boldsymbol{\kappa} \times (\dot {\bf R}_v - {\bf
v}_s)
 \label{magnus}
\end{equation}
acting on a vortex with circulation $\boldsymbol{ \kappa} =\hat {\bf z}h/m$, moving with respect to a superfluid component with density $\rho_s$ and local velocity ${\bf v}_s({\bf R}_v)$, taken at the position ${\bf R}_v(t)$ of the vortex. Thus the Magnus force is also strictly local in space and time, and takes the same form in the HVI and TS equations.

This form for the Magnus force is believed to be exact \cite{Haldane85,Ao93,ThoulessI}; the most convincing proofs of this rely on topological arguments. The Magnus force is therefore independent of sample geometry and of any details of the motion of the vortex.

\vspace{2mm}

{\bf (iii) The quasiparticle force:}  In the HVI equations, the quasiparticle
force ${\bf f}_{qp}^{(cl)}$  is a classical force - it arises from a treatment which treats the vortex as a classical object. It can be written in the form
\begin{equation}
{\bf f}_{qp}^{(cl)} \;=\; D_o ({\bf v}_n - \dot {\bf R}_v) + D_o^{\prime}
\hat {\bf z} \times ({\bf v}_n - \dot {\bf R}_v)
 \label{f-QP}
\end{equation}
in terms of a normal fluid with local velocity ${\bf
v}_n({\bf R}_v)$ and normal fluid density $\rho_n$. The diffusion coefficients $D_o(T)$, $D_o^{\prime}(T)$ (here we use the original notation of HVI) depend on the temperature
$T$; the classic discussion of Iordanskii \cite{Iordanskii64} yields
\begin{equation}
D_o^{\prime}(T) = - \kappa \rho_n(T)
 \label{iordF}
\end{equation}

In the TS equation, on the other hand, the quasiparticle force is history-dependent, ie., it depends on the previous path of the vortex. It takes into account the time-retarded and spatially non-local interaction between thermal quasiparticles and the vortex, and is conveniently decomposed into longitudinal and transverse components, which we write as
\begin{widetext}
\begin{equation}
\vc{f}_{\mathrm{qp}}(t)\;\;\;=\;\;\; \int_{-\infty}^t\ud s \; \gamma_{\parallel}(t-s;T)\; ({\bf v}_n({\bf R}_v)-\dot{\bf R}_v(s)) \;\;+\;\; \int_{-\infty}^t\ud s \; \gamma_{\perp}(t-s;T)\; [\vc{\hat{z}}\times({\bf v}_n({\bf R}_v)-\dot{\bf R}_v(s))],
 \label{fqp-1}
\end{equation}
\end{widetext}
where we note that ${\bf v}_n({\bf R}_v(s))$ is taken to be the local velocity of the normal fluid at the position of the vortex at the time $s$ (rather than the later time $t$ at which the force is acting on the vortex, when it is at position ${\bf R}_v(t)$).

The functions $\gamma_{\parallel}(t;T)$ and $\gamma_{\perp}(t;T)$ are memory kernels. These were found by Thompson and Stamp \cite{TS1,TS2}, and are most easily described in terms of their Fourier transforms - we will define Fourier transforms in this paper by $f(t)=\int_{-\infty}^\infty\ud\omega/(2\pi) f(\omega)e^{i\omega t}$. The memory kernels then take the form
\begin{align}
\tilde{\gamma}_{\parallel}(\omega;T)&\;\;=\;\;D_0(T)\; g_\parallel\left(\frac{\omega}{\omega_T}\right)\\
\tilde{\gamma}_{\perp}(\omega;T)&\;\;=\;\;-\kappa\rho_n(T)\;+\;\frac{1}{2}D_0(T)\;
\frac{\omega_T}{\Lambda_o}\;g_\perp\left(\frac{\omega}{\omega_T}\right)
\end{align}
where $\rho_n$ is again the normal fluid density, and where $D_0(T)$ is a temperature-dependent coefficient given by
\begin{equation}
D_0(T)\;=\;\frac{3m_0\pi\zeta(4)\Lambda_o}{L_z}\left(\frac{k_BT}{\hbar\Lambda_o}\right)^4\;=\;
\frac{3m_0\pi\zeta(4)}{L_z}\frac{\omega_{T}^4}{\Lambda_o^3}.\;\;\;\;
\end{equation}
Here $\zeta(x)$ is the Riemann zeta function. In these equations we have incorporated the frequency dependence of the memory kernels into functions $g_{\parallel}, g_{\perp}$. In addition to  the dimensionless frequency variable $\tilde{\omega}\equiv\omega/\omega_T$, we will also define the dimensionless energy variable $\tilde{\epsilon}_k = \epsilon_k/k_BT$, and the Bose-Einstein function $n_k\equiv(e^{\tilde{\epsilon}_k}-1)^{-1}$. Then the longitudinal function $g_{\parallel}$ takes the form
\begin{equation}
g_\parallel(\tilde{\omega})=\frac{1}{24\zeta(4)|\tilde{\omega}|}
\int_0^\infty\ud \tilde{\epsilon}_k \left(n_k-n_{k+|\tilde\omega|}\right)\frac{\tilde{\epsilon}_k^3
\left(\tilde{\epsilon}_k+\frac{|\tilde\omega|}{2}\right)^2}{\tilde{\epsilon}_k+|\tilde\omega|}
 \label{g-par}
\end{equation}
while the transverse function $g_{\perp}$ involves the same combination of thermal factors, but has a different dependence on
$\tilde{\omega}\equiv\omega/\omega_T$ and $\tilde{\epsilon}_k$; one finds
\begin{equation}
g_\perp(\tilde{\omega})=-\frac{i\mathrm{sgn}(\tilde\omega)}{48\zeta(4)}\int_0^\infty\ud \tilde{\epsilon}_k \left(n_k-n_{k+|\tilde\omega|}\right)\tilde{\epsilon}_k^2\left(\tilde{\epsilon}_k+
\frac{|\tilde\omega|}{2}\right)^2.
 \label{g-perp}
\end{equation}
We notice here that the transverse memory kernel $g_\perp(\tilde{\omega})$ appears at higher order in the variable $\omega_T/\Lambda_o$ than for all other terms in ${\bf f}_{qp}$. This suggests an approximation which we will use in the rest of this paper, which is to drop the frequency dependence of $\tilde{\gamma}_{\perp}(\omega;T)$, ie., we will henceforth write
\begin{equation}
\gamma_{\perp}(t;T) \;\; \rightarrow \;\; -\kappa \rho_n(T) \delta(t)
 \label{gamPerp'}
\end{equation}
so that the transverse term in (\ref{fqp-1}) is now a delta-function in time. The approximation of dropping the non-local part of $\gamma_{\perp}(t;T)$ is obviously very good in the low $T$ regime - a quantitative discussion of the corrections to (\ref{gamPerp'}) coming from this non-local part is given in ref \cite{TS1} (where it is called $d_{\perp}(t;T)$).

The resulting equation for ${\bf f}_{qp}(t)$ is then much simpler. All non-local "memory" effects in ${\bf f}_{qp}(t)$ now come entirely from the longitudinal drag force on the vortex. Thus
we can now write the quasiparticle force in the form that we will use it for the rest of this paper, as
\begin{align}
\vc{f}_{\mathrm{qp}}(t)\;&=\; \int_{-\infty}^t\ud s \; \gamma_{\parallel}(t-s;T)\; ({\bf v}_n-\dot{\bf R}_v(s)) \nonumber \\
& \qquad\qquad + \; \gamma_{\perp}(T)\; [\vc{\hat{z}}\times({\bf v}_n-\dot{\bf R}_v(t))],
 \label{fqp-1'}
\end{align}
with $\gamma_{\perp}(T) =  - \kappa \rho_n(T)$. Thus, in this approximation, the difference between the HVI quasiparticle force and the TS force is entirely in the non-local part of $\gamma_{\parallel}(t-s;T)$ contained in the Fourier transform of $g_\parallel(\tilde{\omega})$.

\vspace{2mm}

{\bf (iv) The fluctuation force:} This force must arise in any quantum Langevin equation, and is due to a combination of quantum and thermal noise. In general it is defined by its time correlator
\begin{equation}
\left\langle f_\mathrm{fl}^a(t)f_\mathrm{fl}^b(s)\right\rangle\;\equiv\; \chi^{ab}(t-s;T)
 \label{chi-ab}
\end{equation}
The behaviour of the correlator is again most easily seen by looking
at its Fourier transform; this is found to be
\begin{widetext}
\begin{align}
\tilde{\chi}^{ab}(\omega,T)& \;\;= \;\; \tilde{\chi_\parallel}\left(\omega;T\right)\delta^{ab}\;+\;\tilde{\chi}_\perp
\left(\omega;T\right)\varepsilon^{ab} \nonumber\\
& \;\;\equiv \;\; \chi_0(T)\left[\eta_\parallel\left(\frac{\omega}{\omega_T}\right)\delta^{ab}\;+\;
\frac{\omega_T}{2\Lambda_o}\eta_\perp\left(\frac{\omega}{\omega_T}\right)\epsilon^{ab}\right]
 \label{chi-fl}
\end{align}
where we have defined "friction" coefficients
\begin{align}
\eta_\parallel(\tilde{\omega})&\;=\;\frac{15}{8\pi^4}\int_0^\infinity\ud \tilde{\epsilon}_k \left\{\frac{\tilde{\epsilon}_k^6}{(\tilde{\epsilon}_k +|\tilde{\omega}|)^2}n_k(1+n_{k+|\tilde{\omega}|})+(\tilde{\epsilon}_k +|\tilde{\omega}|)^2 \tilde{\epsilon}_k^2 \; n_{k+|\tilde{\omega}|}(1+n_k)\right\} \\
\eta_\perp(\tilde{\omega})&\;=\;\frac{15\tilde{\omega}^4}{8\pi}\int_0^\infinity\ud \tilde{\epsilon}_k \; (\tilde{\epsilon}_k +|\tilde{\omega}|)^2 (2\tilde{\epsilon}_k +|\tilde{\omega}|) \tilde{\epsilon}_k \;  n_k(1+n_{k+|\tilde{\omega}|})
\end{align}

\end{widetext}
and where the frequency-independent part of the correlator is written as
\begin{equation}
\chi_0(T)\;=\; {\hbar \over \pi}\omega_TD_0(T) \; \equiv \; {k_BT \over \pi} D_0(T)
 \label{eq:chi0}
\end{equation}
Note that if the fluctuations could be treated as Markovian, this last term would be the only term in the noise correlator. However it is incorrect to do this in general. Moreover it is actually incorrect to drop the noise term under any circumstances - the presence of a frictional drag term in the quasiparticle force necessarily requires a noise term. This shows that there is a basic inconsistency in the HVI equation, which contain no noise term. If one inserts it by hand, then for the HVI equations it would take precisely the Markovian form just given in (\ref{eq:chi0}). The non-Markovian part of $\chi^{ab}(t-s;T)$, coming from the frequency dependence of the friction coefficients in $\tilde{\chi}^{ab}(\omega,T)$, is directly associated with the retarded part of $\gamma_{\parallel}(\omega,T)$.

There is an interesting subtlety here. One might think that it is inconsistent to keep the retarded part of the transverse fluctuation correlator, if at the same time we drop the retarded part of the transverse drag term. To check this, we expand $\boldsymbol{S}_{vv}(\omega)=\vc{H}(\omega)\vc{X}(\omega)\vc{H}^T(-\omega)$ in $\omega_T/\Lambda_0$ to get:
\begin{widetext}
\begin{align}
S_\vc{vv}^\parallel(\omega)\;\;\sim&\;\;\frac{\chi_\parallel(\omega)}{(\rho\kappa)^2}+\frac{1}{(\rho\kappa)^3}\left[-\chi_\parallel(\omega)\left(\gamma_\perp(\omega)+\gamma_\perp(-\omega)\right)+\chi_\perp(\omega)\left(\gamma_\perp(\omega)+\gamma_\perp(-\omega)\right)\right]+\ldots\\
=&\;\; \mathcal{O}\left[\left(\frac{\omega_T}{\Lambda_0}\right)^3\right]+\mathcal{O}\left[\left(\frac{\omega_T}{\Lambda_0}\right)^8\right]+\ldots
\end{align}
\end{widetext}
for the longitudinal correlator, and 
\begin{align}
S_\vc{vv}^\perp(\omega)\;\;\sim&\;\;\frac{\chi_\perp(\omega)}{(\rho\kappa)^2}+\frac{\chi_\parallel(\omega)}{(\rho\kappa)^3}\left[\gamma_\parallel(-\omega)-\gamma_\parallel(\omega)\right]+\ldots\\
=&\;\;\mathcal{O}\left[\left(\frac{\omega_T}{\Lambda_0}\right)^3\right]+\mathcal{O}\left[\left(\frac{\omega_T}{\Lambda_0}\right)^7\right]+\ldots\\
\end{align}
and we see that $\gamma_\perp$ doesn't appear in the expression for $S^\perp_{\vc{vv}}$ until order $\mathcal{O}\left[\left(\frac{\omega_T}{\Lambda_0}\right)^9\right]$.

\subsection{Equations of motion: simplified form}
 \label{sec:Eom-item}

We may now summarize what we have found, in the form of equations of motion including inertial forces, quasiparticle forces, fluctuation forces, and any external applied forces.
The equations of motion are just that in (\ref{eom2}), with the Magnus and inertial forces given by the usual local expressions, with the non-local quasiparticle force given by (\ref{fqp-1}), and with a fluctuation noise force having a correlator given by (\ref{chi-fl}).

However, in the approximation where we keep only the local "Iordanskii" part of the transverse drag function $\gamma_{\perp}(\omega, T)$, things greatly simplify: the quasiparticle force can then be written as (\ref{fqp-1'}). If we now add together the Iordanskii force, of magnitude $-\kappa \rho_n(T)$, and the Magnus force, of size $-\kappa \rho_s(T)$, we can write the equations of motion in the simple form
\begin{equation}\label{eq:om}
M_v\ddot{\bf R}_v({t})+ {\bf f}_{qp}(t) )\;=\;\tilde{\bf F}(t),
\end{equation}
where the total "driving" force $\tilde{\bf F}(t) = {\bf F}_{ac}(t) + {\bf f}_{fl}(t)$ is now the sum of any external force and the internal fluctuation force; and where
\begin{align}
{\bf f}_{qp}(t) \;=\;  \int_{-\infty}^{{t}}\ud s \; \boldsymbol{\sigma}_0 \; \gamma_{\parallel}({t}-s) \dot{\bf R}_v(s)
\;+\; i \boldsymbol{\sigma}_2  \kappa \rho \dot{\bf R}_v(t) \;\;
 \label{fqp-2}
\end{align}
is the quasiparticle force, with a purely local transverse component summing the Magnus and Iordanskii terms. Here we introduce a notation in which the usual Pauli matrices $\boldsymbol{\sigma}_j$ operate on vectors in the 2-dimensional plane; $\boldsymbol{\sigma}_0$ is the $2 \times 2$ identity matrix, and
\begin{equation}
-i\boldsymbol{\sigma}_2 \equiv \begin{pmatrix}0&-1\\1&0\end{pmatrix}.
\end{equation}

In the rest of this paper we will be concerned with the solution of eqtn. (\ref{eq:om}) in different regimes, including those that may be accessible in experiment.


\section{Formal Solution of Equations of Motion}\label{sec:Derivation}
 \label{sec:derivation}


To give a formal solution of the TS equation, we stay in two dimensions - we briefly comment on the 3-dimensional case later in the paper. We therefore consider the motion of a vortex in two dimensions with no background fluid flow (ie., in the rest frame of the superfluid). We consider the case where the vortex has started its motion from rest at $\vc{r}=0$ a long time in the past; the results are easily modified to deal with other initial conditions.

We note also that when there is a background flow of either (or both)  of the superfluid or normal fluid components, we can treat the effect of this background flow as an external force $\vc{f}_{\textrm{bf}}$ acting on the vortex given by
\begin{equation}
\vc{f}_{\textrm{bf}}\;\;=\;\; i \boldsymbol{\sigma}_2 \kappa {\bf J} + \int_{-\infty}^t\ud s\gamma_{\parallel}(t-s)\vc{v}_n(s)
 \label{fbf}
\end{equation}
where
\begin{equation}
{\bf J}({\bf R}_v) \;\equiv\; \rho\bar{\vc{v}}\;\;=\;\;(\rho_s\vc{v}_s+\rho_n\vc{v}_n)
 \label{J+v}
\end{equation}
is the total mass current at the position of the vortex.

The equations of motion can be solved for the vortex position ${\bf R}_v(t)$ and velocity ${\bf V}_v(t) \equiv {\bf \dot{R}}_v(t)$, using position and velocity Green's functions $\vc{G}(t)$ and $\vc{H}(t)$, defined as
\begin{align}
{\bf R}_v(t)& \;=\;\int_{-\infty}^\infty\ud s\vc{G}(t-s)\cdot\vc{f}(s)\\
{\bf V}_v(t)& \;=\;\int_{-\infty}^\infty\ud s\vc{H}(t-s)\cdot\vc{f}(s).
 \label{rv-sol}
\end{align}
where these two Green functions are related by
\begin{equation}
\vc{H}(t)= {d\vc{G}(t) \over dt}
 \label{GH-rel}
\end{equation}
and we have written $d\vc{G}(t)/dt$ as $\vc{H}(t)$ to make it easier to read. Their Fourier transforms are given from (\ref{eq:om}) by
\begin{align}
\vc{G}(\omega)&\;\;=\;\;\frac{(\omega M_v-i{\tilde{\gamma}_R(\omega)})\boldsymbol{\sigma}_0-{{\rho\kappa}}\boldsymbol{\sigma}_2} {\omega\left[{{\rho^2\kappa^2}}-\left(\omega M_v-i{\tilde{\gamma}_R(\omega)}\right)^2\right]}\label{eq:fullG}\\
\vc{H}(\omega)&\;\;=\;\; i \omega \vc{G}(\omega)\label{eq:fullH}
\end{align}
where $\gamma_R(\omega)$ is the Fourier transform of the retarded forms of the kernel $\gamma_{\parallel}(t)$, viz.,
\begin{equation}
\gamma_R(\omega)=\int_0^\infty\ud t\gamma_{\parallel}(t)e^{-i\omega t}.
\end{equation}

The eqtns. (\ref{rv-sol}) directly give the response of the vortex motion to a deterministic force; in particular, the response to a periodic force ${\vc{f}}(t)=\vc{f}_\omega e^{i\omega t}$ is just
\begin{align}
{\bf R}_v(\omega) &\;=\; \vc{G}(\omega)\cdot \vc{f}_\omega\\   {\bf V}_v(\omega) &\;=\;\vc{H}(\omega)\cdot \vc{f}_\omega.
 \label{r-per}
\end{align}
where we note that the force acting on the vortex will, according to (\ref{eq:om}), be just the total driving force $\tilde{\bf F}(t)$.

Adding the noise terms to the equations of motion means we no longer have a deterministic response - the vortex variables become stochastic. We can then describe the response to the fluctuation force by auto-correlation functions, defined as
\begin{align}
\vc{S}_{\vc{RR}}(t-t') \;&=\; \left\langle {\bf R}_v(t)\otimes {\bf R}_v(t')\right\rangle  \\
\vc{S}_{\vc{vv}}(t-t') \;&=\; \left\langle {\bf V}_v(t)\otimes {\bf V}_v(t')\right\rangle
\end{align}
where $\otimes$ denotes the outer product, so that $(\vc{x}\otimes\vc{y})_{ab}=x_ay_b$. We then have the stochastic equations of motion
\begin{widetext}
\begin{align}\label{eq:autocor}
\vc{S}_{\vc{RR}}(t,t')  &\;\;=\;\; {\bf R}_v(t) \otimes {\bf R}_v(t') \;+\; \int_{-\infty}^\infty\ud s\int_{-\infty}^{\infty}\ud s'\vc{G}(t-s)\cdot\left\langle \vc{f}(s)\otimes\vc{f}(s')\right\rangle\cdot\vc{G}^T(t'-s')\\
\vc{S}_{\vc{vv}}(t,t')  &\;\;=\;\; {\bf V}_v(t) \otimes {\bf V}_v(t') \;+\; \int_{-\infty}^\infty\ud s\int_{-\infty}^{\infty}\ud s'\vc{H}(t-s)\cdot\left\langle \vc{f}(s)\otimes\vc{f}(s')\right\rangle\cdot\vc{H}^T(t'-s').
\end{align}
\end{widetext}
Here the first terms ${\bf R}_v(t) \otimes {\bf R}_v(t')$ and ${\bf V}_v(t) \otimes {\bf V}_v(t')$ on the right-hand side of (\ref{eq:autocor}) are given directly from the solution of the deterministic equations (\ref{rv-sol}); and the force ${\bf f}(t)$ is whatever total driving force happens to be acting on the vortex.

If the driving force were purely stochastic, ie., there were no deterministic forces at all, and $\left\langle \vc{f}(t)\right\rangle=0$, then only the second terms on the right-hand side of (\ref{eq:autocor}) would remain. In this case the motion of the vortex is just given by
\begin{align}
\vc{S}_{\vc{RR}}(\omega)& \;=\; \vc{G}(\omega)\cdot \vc{X}(\omega)\cdot \vc{G}^{T}(-\omega)\label{eq:vvacfsp}\\
\vc{S}_{\vc{vv}}(\omega)& \;=\; \vc{H}(\omega)\cdot \vc{X}(\omega)\cdot \vc{H}^{T}(-\omega)
\end{align}
where
\begin{equation}
\vc{X}(t)=\chi_\parallel(t)\boldsymbol{\sigma}_0-i\chi_\perp(t)\boldsymbol{\sigma}_2.
\end{equation}
is just the Fourier transform of the fluctuation correlator in (\ref{chi-fl}), and where $\vc{S}_{\vc{rr}}(\omega), \vc{S}_{\vc{vv}}(\omega)$ are just the Fourier transforms of the time autocorrelators defined above.

We see that the vortex dynamics is now much more complex than that given by the simple HVI equations (\ref{eom1}). In the following sections we progressively add terms to the HVI equation, solving for each case, to see how the different additional terms affect the vortex dynamics.


\section{Classical Regime: HVI Equations plus noise and inertial terms}
\label{sec:HVI}


In this section we will ignore all the memory effects that exist in the general TS equation, and simply look at how the HVI equations is affected by the addition of inertial terms and uncorrelated (Markovian) white noise. These results will then only be strictly valid in the high temperature/low frequency classical regime, where we can use an entirely local description of the vortex dynamics.

\subsection{Solution for bare HVI equations}
 \label{sec:HVI pure}

As a check we first solve the equations of motion without any additional terms, ie., we solve the original HVI equation.
In the formalism developed here, this takes the simple form
\begin{equation}
\left\{\gamma_0\boldsymbol{\sigma}_0+i\rho \kappa\boldsymbol{\sigma}_2\right\} {\bf \dot{R}}_v(t)=\vc{f}(t).\label{HVI}
\end{equation}
Where $\gamma_0$ is the zero frequency component of the retarded Fourier transform of the longitudinal friction kernel $\gamma_{\parallel}(t)$. The velocity Green's function can be read from equations \eqref{HVI},
\begin{align}
\vc{H}(t) \;&=\; \delta(t)\left\{\gamma_0\boldsymbol{\sigma}_0+i\Omega\boldsymbol{\sigma}_2\right\}^{-1} \nonumber \\
&=\; \frac{\delta(t)}{\rho^2 \kappa^2 +\gamma_0^2}\left(\gamma_0\boldsymbol{\sigma}_0-i\rho\kappa\boldsymbol{\sigma}_2\right).
\end{align}
Then if the vortex starts from rest at the origin the position Green's function $\vc{G}_0(t)$ is given by
\begin{equation}
\vc{G}(t)=\frac{\Theta(t)}{\rho^2\kappa^2+\gamma_0^2}
\left(\gamma_0\boldsymbol{\sigma}_0-i\rho\kappa\boldsymbol{\sigma}_2\right)
\end{equation}
where $\Theta(t)$ is the unit step function. We see that without inertia, the vortex's velocity responds instantly to an applied force. In the presence of a background fluid flow the vortex velocity is then given by
\begin{align}
{\bf \dot{R}}_v(t) \;=&\;\; \frac{-\gamma_0\rho_s\boldsymbol{\kappa}\times(\vc{v}_s-\vc{v}_n)+
\rho^2\kappa^2\bar{\vc{v}}+\gamma_0^2\vc{v}_n}{\rho^2\kappa^2+\gamma_0^2} \nonumber\\
=& \;\; \bar{\vc{v}}+\frac{\rho_s\gamma_0}{\rho^2\kappa} [\hat{\vc{z}}
\times(\vc{v}_n-\vc{v}_s)] \nonumber\\ &\qquad\qquad +\mathcal{O}\left[(m\omega_T / \rho\kappa)^2, (\omega_T / \omega_\mathrm{c})^6\right]
 \label{eq:avvHVI}
\end{align}
where, as before, $\bar{\vc{v}} = {\bf J}/\rho$ is the velocity associated with the total mass current at the position of the vortex (compare eqtn. (\ref{J+v})). If $(\vc{v}_n-\vc{v}_s) = 0$,
the vortex is carried with the mass flow; otherwise ``mutual friction" operates, and we can rewrite (\ref{eq:avvHVI}) in the standard form \cite{Donnelly91},
\begin{equation}
{\bf \dot{R}}_v \;=\; \vc{v}_s+\alpha\hat{\vc{z}}\times(\vc{v}_n-\vc{v}_s)-\alpha'\hat{\vc{z}}\times
\left[\hat{\vc{z}}\times(\vc{v}_n-\vc{v}_s)\right]
\end{equation}
with the parameters $\alpha$ and $\alpha'$ defined as
\begin{align}
\alpha&=\frac{\rho_s\gamma_0}{\rho^2\kappa}\\
\alpha'&=\frac{\rho_n}{\rho}
\end{align}
so that for low temperatures $\alpha'\sim T^3$ in 2 spatial dimensions.\\

\subsection{HVI equations plus white noise}
 \label{sec:HVI-noise}

In circumstances where $\omega/\omega_T \ll 1$, where $\omega$ is the characteristic frequency of vortex motion, the vortex dynamics is deep into the classical regime. In this case we can treat the noise term as white noise, ie., as a fluctuating force with the Markovian correlator
\begin{equation}
\left\langle {\bf f}_{fl}(t)\otimes {\bf f}_{fl}(s)\right\rangle=\chi_0(T) \delta(t-s)\boldsymbol{\sigma}_0
 \label{fl-0}
\end{equation}
with $\chi_0$ given by equation \eqref{eq:chi0}. In this low-frequency regime we can neglect the transverse part of the noise correlator.

The position-position correlator $\vc{S}_{\vc{RR}}(t,t')$ with initial conditions $\vc{r}(0)=0$ is given by
\begin{equation}
\vc{S}_{\vc{RR}}(t,t')=\chi_0 \frac{\boldsymbol{\sigma}_0}{\rho^2\kappa^2+\gamma_0^2} \min(t,t')
 \label{eq:msdhvi}
\end{equation}
so that, as expected, the vortex mean-squared displacement increases linearly with time. The displacement from a given position is uncorrelated to that position. The position coordinates $x(t)$ and $y(t')$  are uncorrelated to each other. Equation \eqref{eq:msdhvi} gives an effective diffusion coefficient $D$,
\begin{equation}
D=\frac{\chi_0}{(\rho^2\kappa^2+\gamma_0^2)}
\end{equation}
The velocity auto-correlator follows immediately, as
\begin{equation}
\vc{S}_{\vc{vv}}(t,t')= \chi_0\frac{\boldsymbol{\sigma}_0}{\rho^2\kappa^2+\gamma_0^2} \delta(t-t')
 \label{V-delta}
\end{equation}
and the velocities are completely uncorrelated at separate times. The mean squared velocity is then
 \begin{align}
\left\langle V^2_v(t)\right\rangle \;=& \;\vc{S}_{\vc{vv}}(t=t') \nonumber \\
& \sim \; \frac{\chi_0}{\tau_v(\gamma^2+\rho^2\kappa^2)}\boldsymbol{\sigma}_0
 \label{eq:msvhvit}
\end{align}
where we define $\tau_v$ as the time it takes for the vortex velocity to react to the noise force - in the present case where $M_v = 0$, one has $\tau_v = 0$, so that the mean vortex velocity is infinite, and the different components of the velocity vector are uncorrelated even at the same time. We see immediately below that the effect of fluctuation noise will be to smooth the delta-function in (\ref{V-delta}), and give a finite $\tau_v$.

These results are unphysical - an uncorrelated vortex velocity implies instantaneous reaction to noise forces. As noted above, it is inconsistent to have a quasiparticle drag term $\gamma_0$ in the HVI equations without also having a noise force, which in the classical regime we expect to be Markovian. However  the assumption of no inertial mass is also unphysical - any mass will smooth the delta-function in (\ref{V-delta}), and make $\tau_v$ finite.

\subsection{HVI equations plus inertia and white noise.} \label{sec:HVIin}

\begin{figure}
\includegraphics[width=8.6 cm]{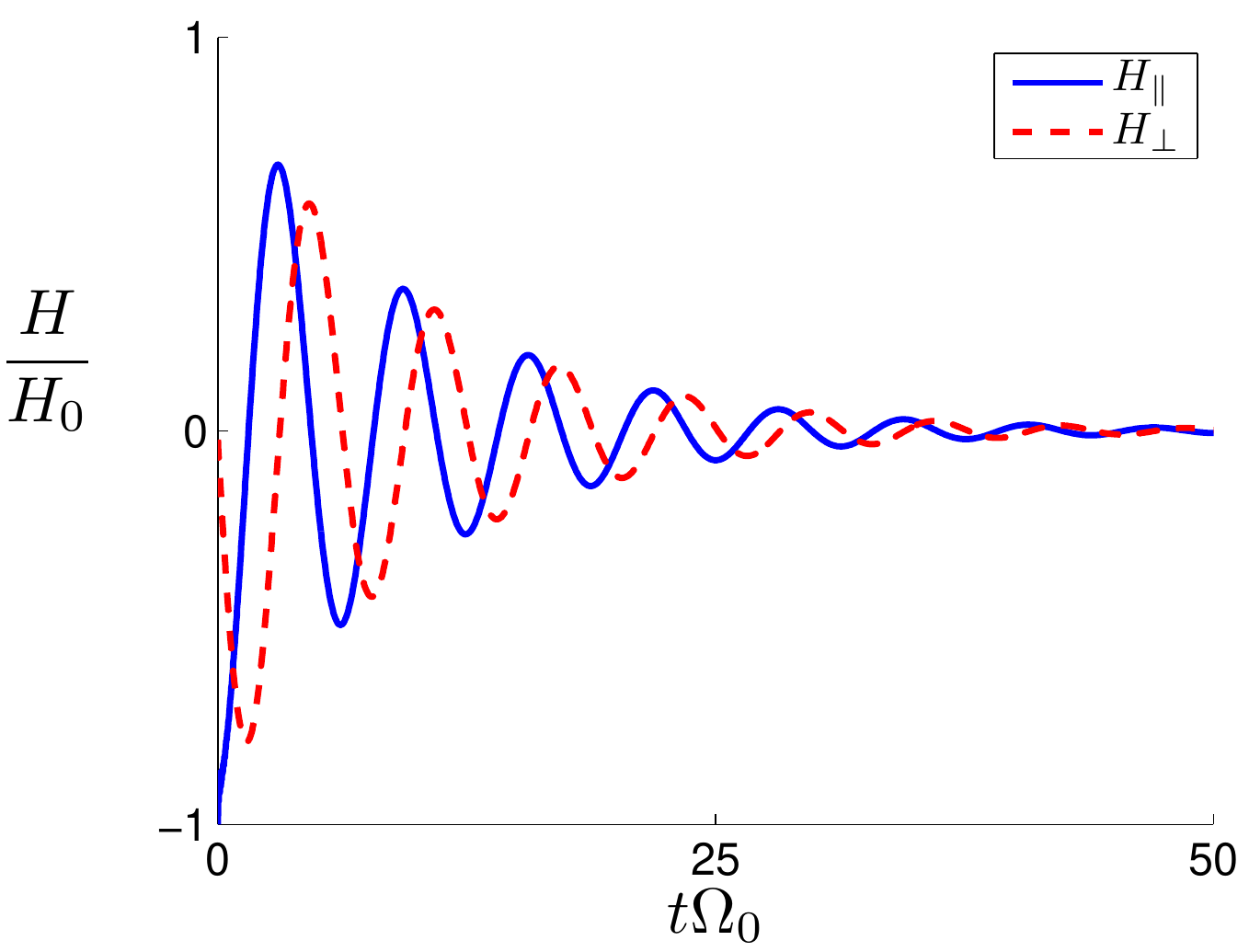}
\caption{The longitudinal (solid line) and transverse (dashed line) components of the velocity Greens function $\vc{H}(t)$ in equation \eqref{eq:HwithM}. The vertical axis is scaled by the value of the longditudnal part of the Green's function immediately after $t=0$, $H_0=|H_\parallel(0+)|$.  The damping time scale and cyclotron frequency are chosen so that $\tau_D\Omega_0=0.1$. \label{fig:Hvst}}
\end{figure}

We consider first the new Green function in the presence of inertia, ignoring the noise term. The inertial force term in the equations of motion gives the $\vc{G}(\omega)$ the form
\begin{align}
\vc{G}(\omega)&=\frac{(\omega M_v-i{{\gamma}_0})\boldsymbol{\sigma}_0-{{\rho\kappa}}\boldsymbol{\sigma}_2} {\omega\left[\rho^2\kappa^2-\left(\omega M_v-i{{\gamma}_0}\right)^2\right]}
 \label{G-in}
\end{align}
and $\vc{H}(\omega) = i\omega \vc{G}(\omega)$. Thus including the inertial mass term has added two poles, at $\omega_\pm=(\pm\Omega+i\gamma_0)/M_v$, to each of these Green functions. Their high frequency behaviour has now changed, so that for $\omega \gg \omega_T$ one has
\begin{align}
\vc{G}(\omega)\sim-\frac{1}{M_v\omega^2}\boldsymbol{\sigma}_0+\frac{\rho\kappa}{\omega^3M_v^2}
\boldsymbol{\sigma}_2
 \label{eq:highfrg}
\end{align}
compared to the inertia free case where $\vc{H}(\omega)=(\gamma_0\boldsymbol{\sigma}_0+i\rho\kappa\boldsymbol{\sigma}_2)^{-1}$, and  $\vc{G}(\omega) = -i\vc{H}(\omega)/\omega$, for all $\omega\ll\omega_\mathrm{c}$. The velocity Green's function in the time domain, shown in figure \ref{fig:Hvst}, is found by contour integration to be
\begin{equation}\label{eq:HwithM}
\vc{H}(t)=-\frac{\Theta(t)e^{-\frac{t}{\tau_D}}}{M_v}[ \boldsymbol{\sigma}_0 \cos\Omega_o t - i \boldsymbol{\sigma}_2 \sin\Omega_o t].
\end{equation}
where
\begin{align}
\Omega_o \;=&\; \rho \kappa/M_v \nonumber \\
\tau_D\;=&\;\frac{M_v}{\gamma_0}
\end{align}

\begin{figure}
\includegraphics[width=8.6 cm]{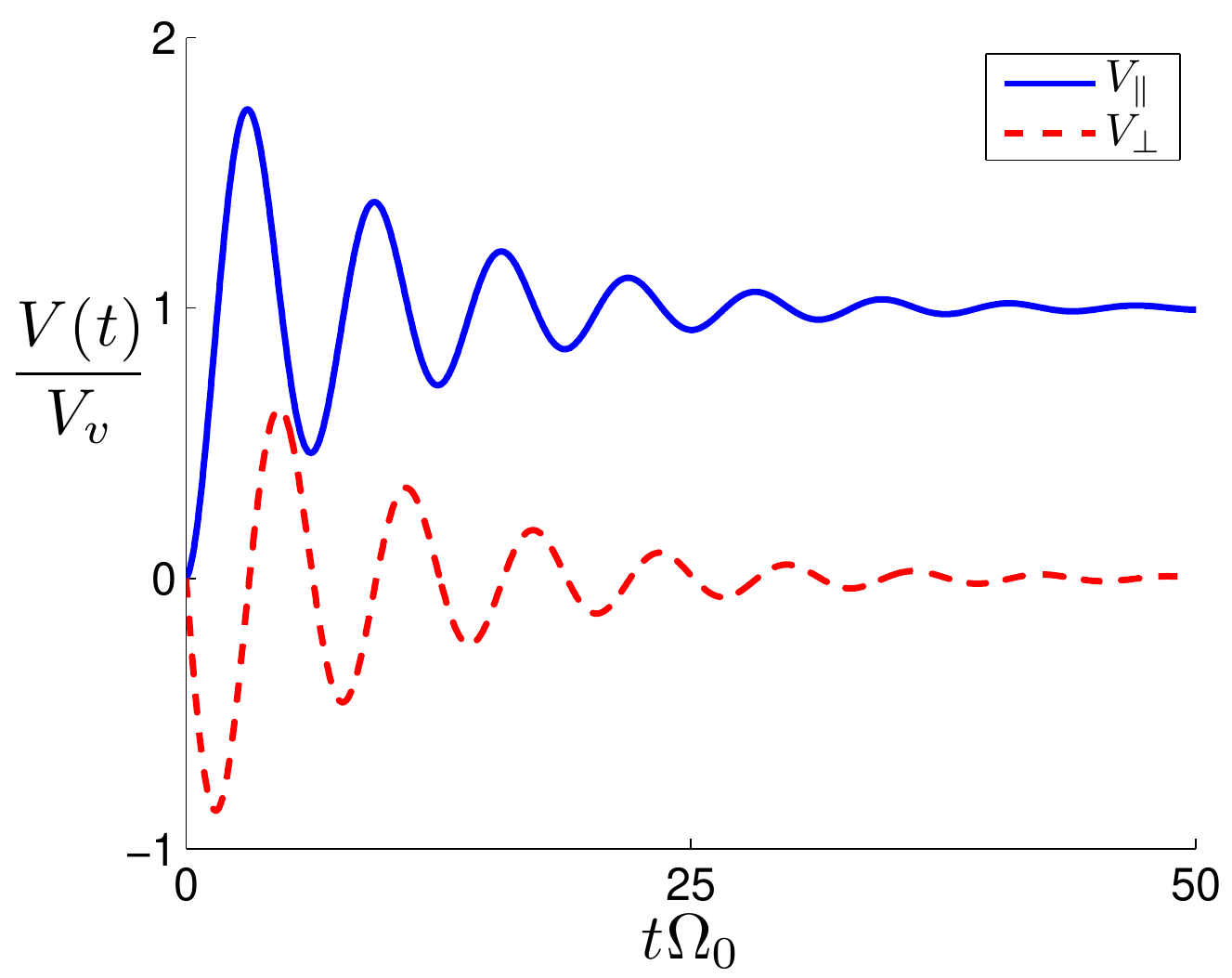}

\caption{\label{fig:vvstvL}The solution to the HVI equations with inertia when the vortex is released from rest into a background superflow. The components of the vortex velocity which are parallel (solid line) and perpendicular (dashed line) to the final velocity $\vc{V}_v$ as a function of time are shown. The damping time scale and cyclotron frequency are chosen so that $\tau_D\Omega_0=0.1$. }
\end{figure}

\begin{figure}
\includegraphics[width=8.6 cm]{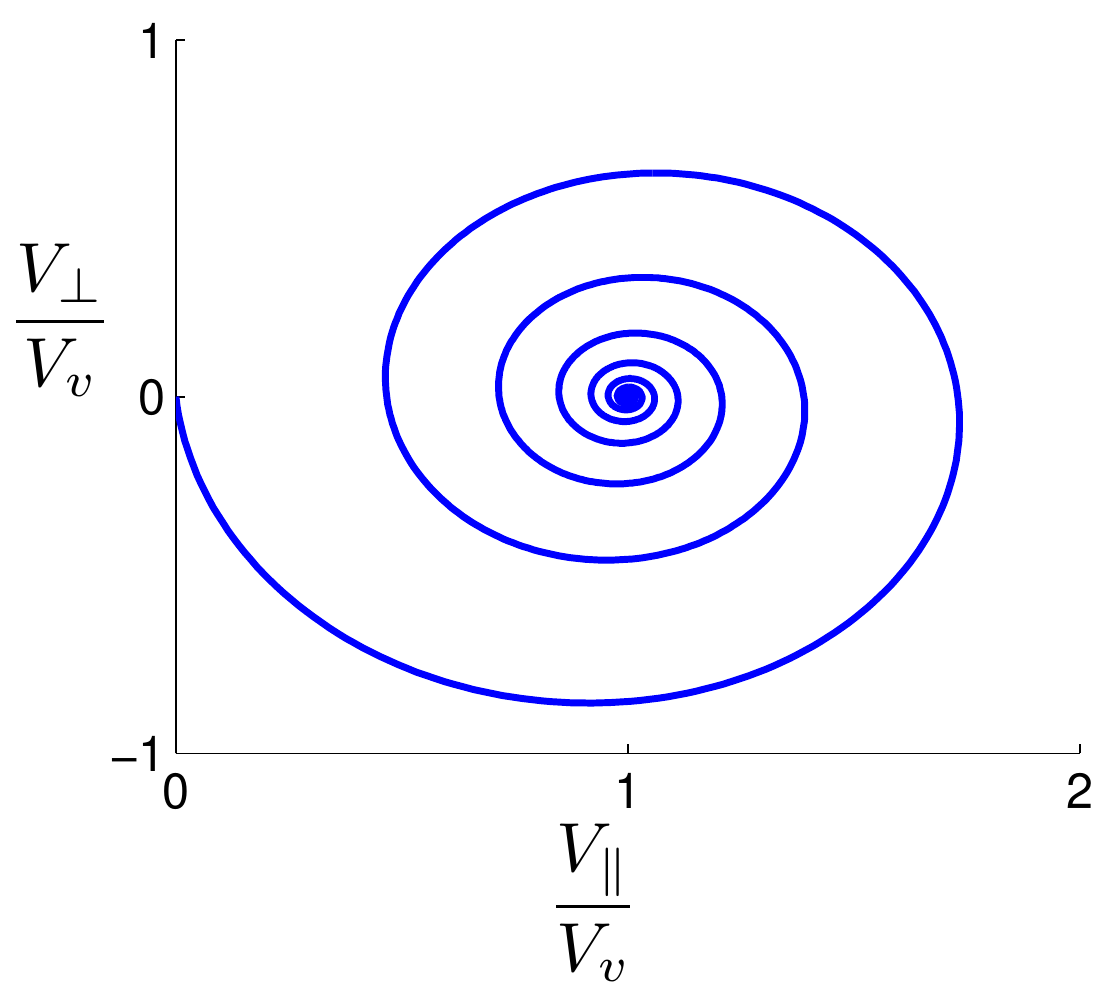}

 \caption{The solution to the HVI equations with inertia when the vortex is released from rest into a bacground superflow. The plot shows the path the vortex's velocity vector traces as it spirals to its final value. \label{fig:vxvyvL}}
\end{figure}

\begin{figure}
\includegraphics[width=8.6cm]{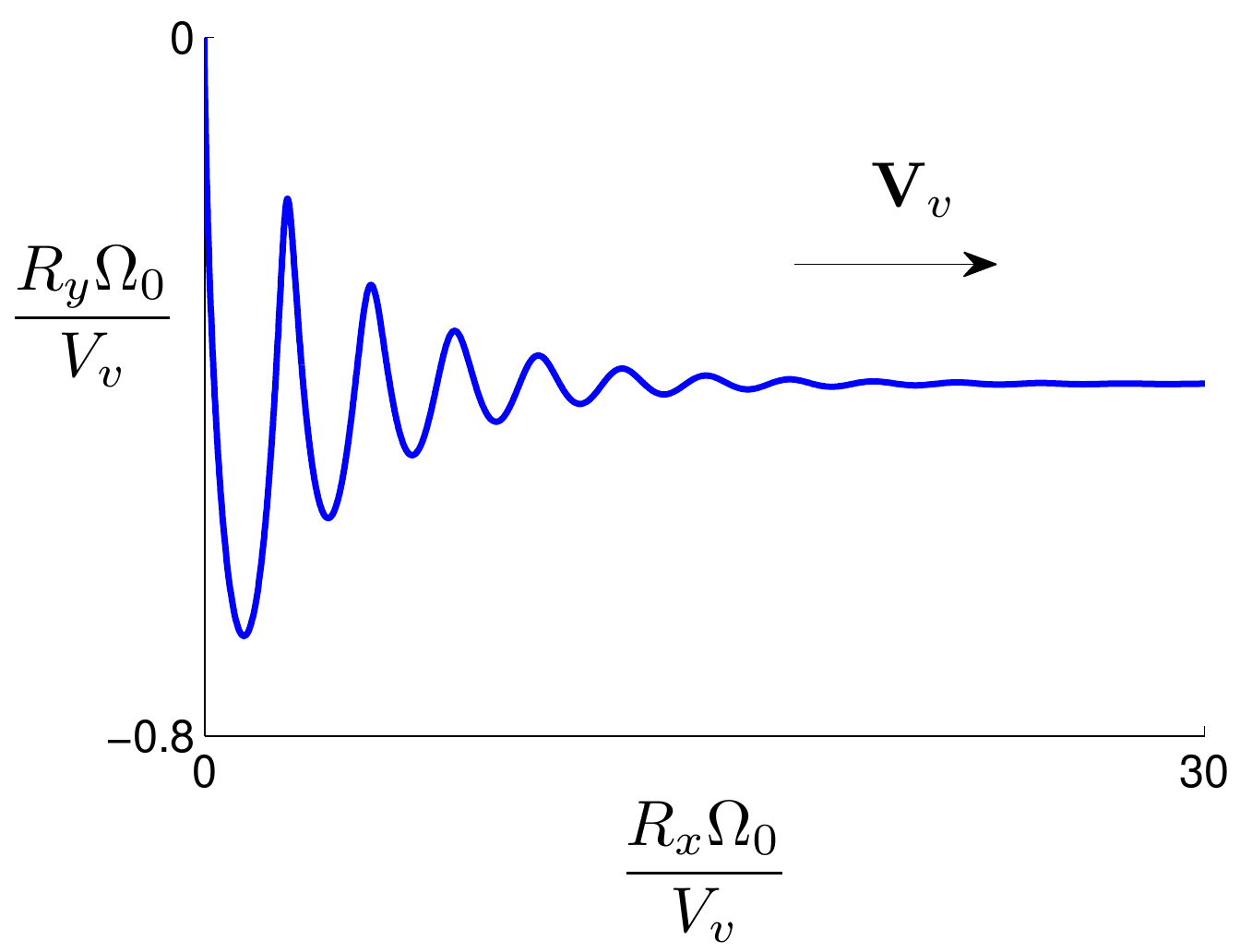}

\caption{\label{fig:xvyvL}The path a of a vortex released from rest into a background superflow. The for large times the the vortex's velocity tends to a value $\vc{V}_v$ which is determined by the balance of the friction force and the Magnus force. The damping time scale and cyclotron frequency are chosen so that $\tau_D\Omega_0=0.1$ and memory effects are not included. }
\end{figure}

Thus the effect of the two poles is to replace an instantaneous response by an underdamped oscillation: the vortex naturally rotates at its cyclotron frequency $\Omega_o$ and this rotation is damped on a time-scale $\tau_D$. The vortex then takes time to react to changes in the background fluid flow, eventually reaching the new steady state vortex velocity ${\bf V}_v$ given in equation \eqref{eq:avvHVI} after oscillations have died down. The response of a vortex  whose motion is governed by the HVI equations to a background fluid flow is shown in Figs. \ref{fig:vvstvL}, \ref{fig:vxvyvL} and \ref{fig:xvyvL}.

\begin{figure}
\includegraphics[width=8.6 cm]{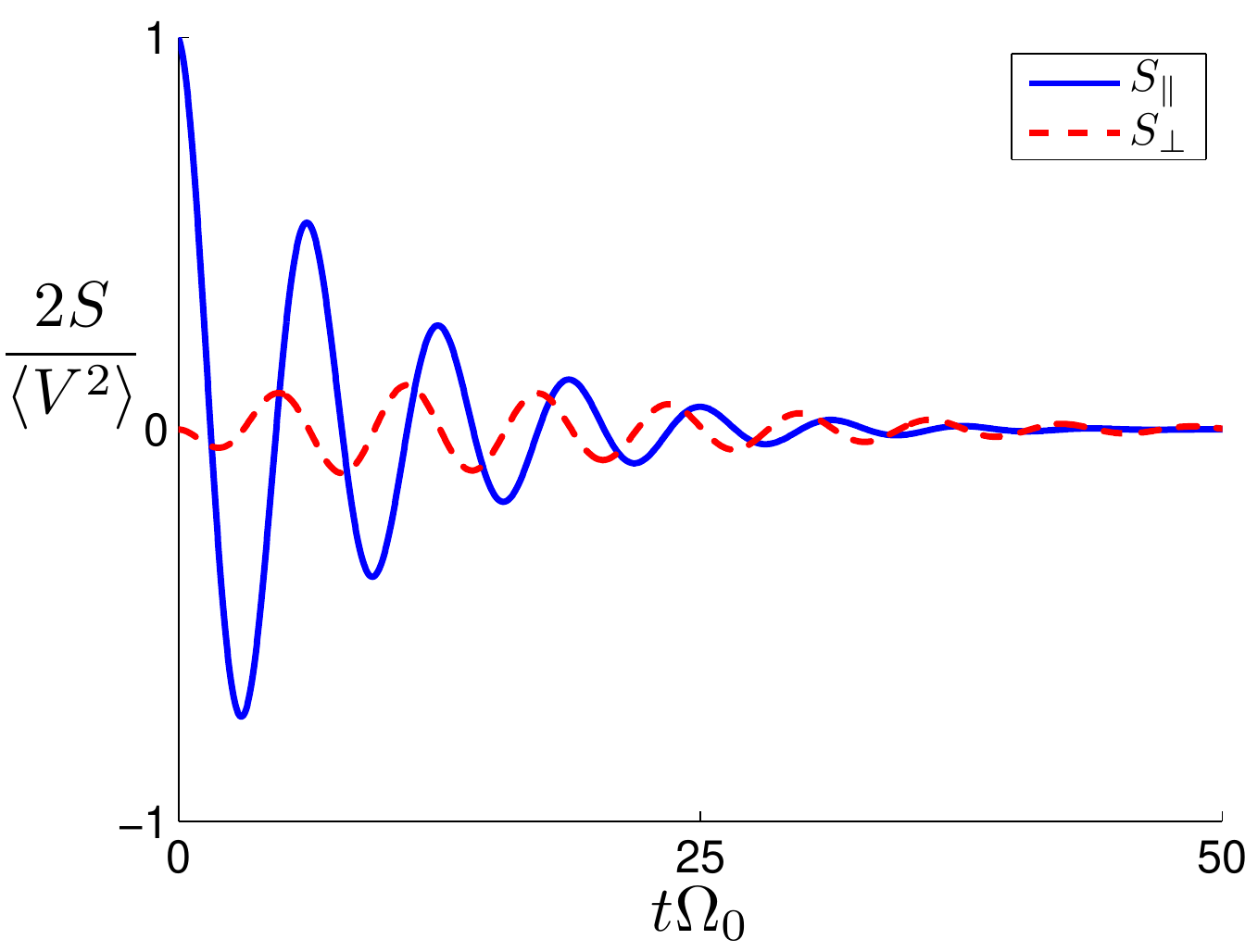}

\caption{\label{fig:Svst}The longitudinal (solid line) and transverse (dashed line) components of the velocity-velocity correlation function $\vc{S}_{\vc{vv}}(t)$ when the effects of inertia and the fluctuation force are considered but the memory forces are ignored. The damping time scale and cyclotron frequency are chosen so that $\tau_D\Omega_0=0.1$.  }
\end{figure}


Now we add white noise as before. The velocity auto-correlator $\vc{S}_\vc{vv}(\omega)$ is still analytic at $\omega=0$. From \eqref{eq:vvacfsp} and \eqref{eq:HwithM} we see that correlations between different velocity components will now die down like $e^{-2t/\tau_D}$, and oscillate at the cyclotron frequency. The time dependence of the velocity-velocity correlation functions is shown in figure \ref{fig:Svst}. Because the high-frequency forms of  $\vc{G}(\omega)$ and  $\vc{H}(\omega)$ are changed by inertia, both $\vc{S}_{\vc{RR}}(\omega)$ and $\vc{S}_{\vc{vv}}(\omega)$ are also changed:
\begin{align}
\vc{S}_{\vc{RR}}(\omega)&\;\sim\; \frac{\chi_0}{M_v^2\omega^4}\boldsymbol{\sigma}_0-\frac{\chi_0\rho\kappa}{M_v^2\omega^5}
\boldsymbol{\sigma}_2 \qquad (\omega \gg \omega_T)
\label{eq:srrhif}\\
\vc{S}_{\vc{vv}}(\omega)&\;\sim\;\frac{\chi_0}{M_v^2\omega^2}\boldsymbol{\sigma}_0+
\frac{\chi_0\rho\kappa}{M_v^2\omega^3}\boldsymbol{\sigma}_2 \qquad (\omega \gg \omega_T)
\label{eq:svvhif}
\end{align}
The expression for $\vc{S}_{\vc{vv}}(\omega)$ dies off like $1/\omega^2$ so $\vc{S}_{\vc{vv}}(t)$ will be a continuous function of $t$ at $t=0$. This means we can calculate the mean squared velocity by taking the limit
\begin{align}
\vc{S}_{\vc{vv}}(t=t')  \; \equiv &\; \left\langle\vc{v}(t)\otimes\vc{v}(t)\right\rangle  \nonumber\\
 =& \; \lim_{\delta\rightarrow 0+}\int_{-\infty}^\infty\frac{\ud \omega}{2\pi}\tilde{\vc{S}}_\vc{vv}(\omega)e^{i\omega \delta}
 \label{Svv-0+}
\end{align}
which is evaluated by closing the contour in the upper half-plane and summing the residues from the poles $\omega_\pm$ to give
\begin{equation}
\left\langle\vc{v}(t)\otimes\vc{v}(t)\right\rangle \;=\;\frac{\chi_0}{2M_v\gamma_0}\boldsymbol{\sigma}_0
\label{eq:msv}
\end{equation}
Comparing equations \eqref{eq:msv} to \eqref{eq:msvhvit} we see that the reaction time $\tau_v$ given in in \eqref{eq:msvhvit} must, once we add inertia, become $\tau_v=2M_v\gamma_0/(\gamma_0^2+\rho^2\kappa^2)$. Letting $M_v \rightarrow 0$ then gives us back the unphysical results in  (\ref{V-delta}) and (\ref{eq:msvhvit})

The time evolution of the correlation function $\vc{S}_{\vc{vv}}(t) \equiv \left\langle\vc{v}(t)\otimes\vc{v}(0)\right\rangle$ is illustrated in Fig. \ref{fig:Svst}, for both longitudinal and transverse components, where the oscillation and decay timescales are clear.


\section{TS Equations in the Quantum Regime: Massless case}\label{sec:HVImem}


We now consider what happens when we move from the classical regime into the crossover regime and thence over to the quantum regime. The signal for this is the rising importance of long-time memory terms in the TS equations as $\omega_T/\omega$ decreases. These come into both the drag force $\gamma_{\parallel}$ and the fluctuating noise force - the latter becomes more and more non-Markovian as we lower $\omega_T/\omega$, and so we can no longer treat it as white noise.

It is important to distinguish the effects of inertia in the results that follow. We therefore consider two cases - first, in this section, the massless case, where inertial terms are absent, and then finally, in the next section, the full TS equation.

We begin as before by determining the Green functions in the absence of fluctuations, and then determine the stochastic equation in the presence of these fluctuations. With $M_v = 0$, one has
\begin{equation}
\vc{G}(\omega) \;=\;-\frac{i}{\omega}\left(\frac{\gamma_R(\omega)
\boldsymbol{\sigma}_0-i\rho\kappa\boldsymbol{\sigma}_2} {\rho^2\kappa^2+ \gamma_R^2(\omega)}\right)
\label{eq:gmemnoM}
\end{equation}
and $\vc{H}(\omega) = i\omega \vc{G}(\omega)$ as before. The
analytic structure of both Green's functions is changed as they
inherit the branch cuts present in the retarded friction kernel
$\gamma_R(\omega)$, in addition to the pole already present in
$\vc{G}(\omega)$ in the inertia-free memoryless case.

\begin{figure}
\includegraphics[width=8.6 cm]{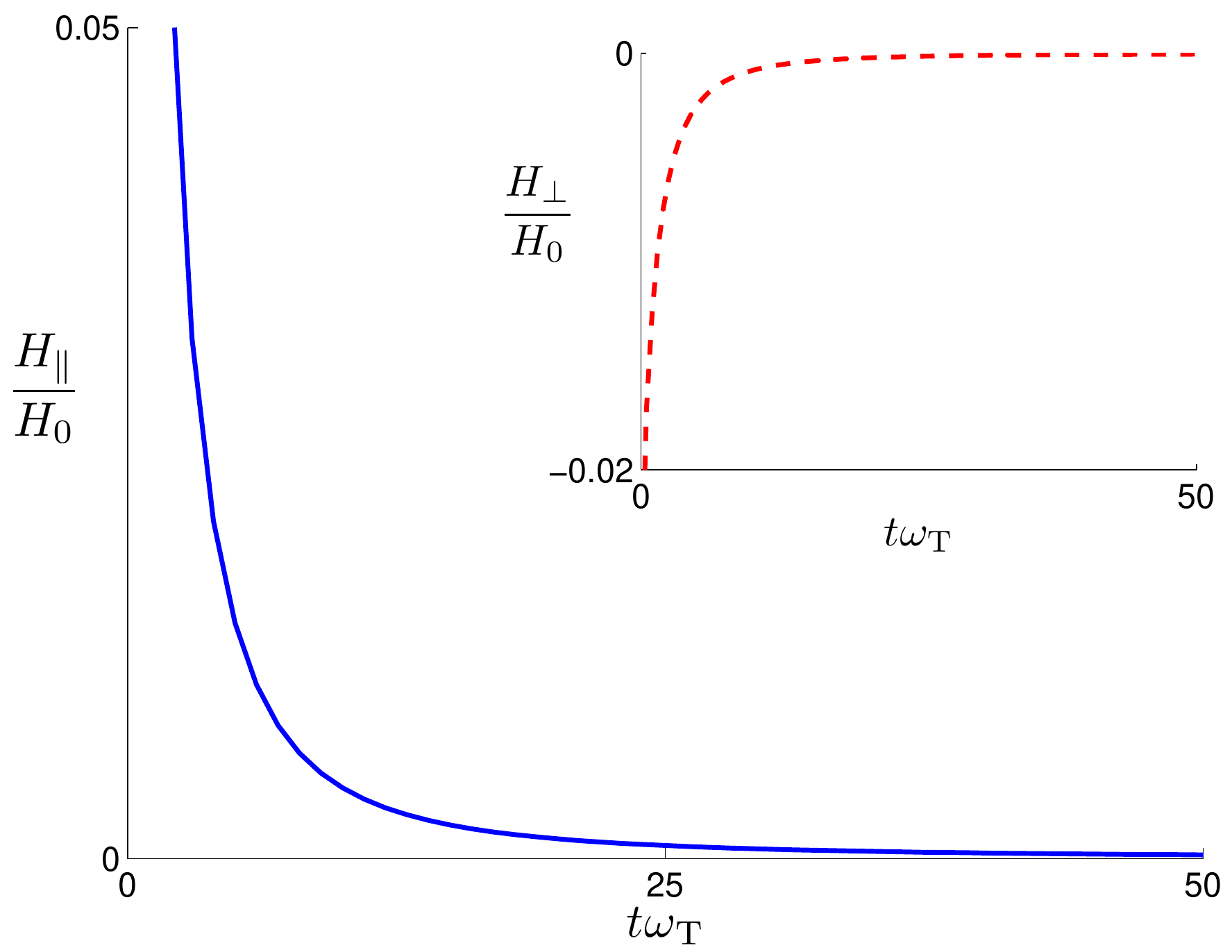}

\caption{\label{fig:Hllmem}The longitudinal (main figure) and transverse (inset) components of the vortex velocity Greens function $\vc{H}(t)$ when the effects of memory are included and inertia is ignored. The vertical axis is scaled by the value of the longitudinal part of the Green's function immediately after $t=0$, $H_0=|H_\parallel(0+)|$. The ratio of the zero frequency friction and magnus coefficient is $\gamma_0/\rho\kappa=0.1$}
\end{figure}

\begin{figure}
\includegraphics[width=8.6 cm]{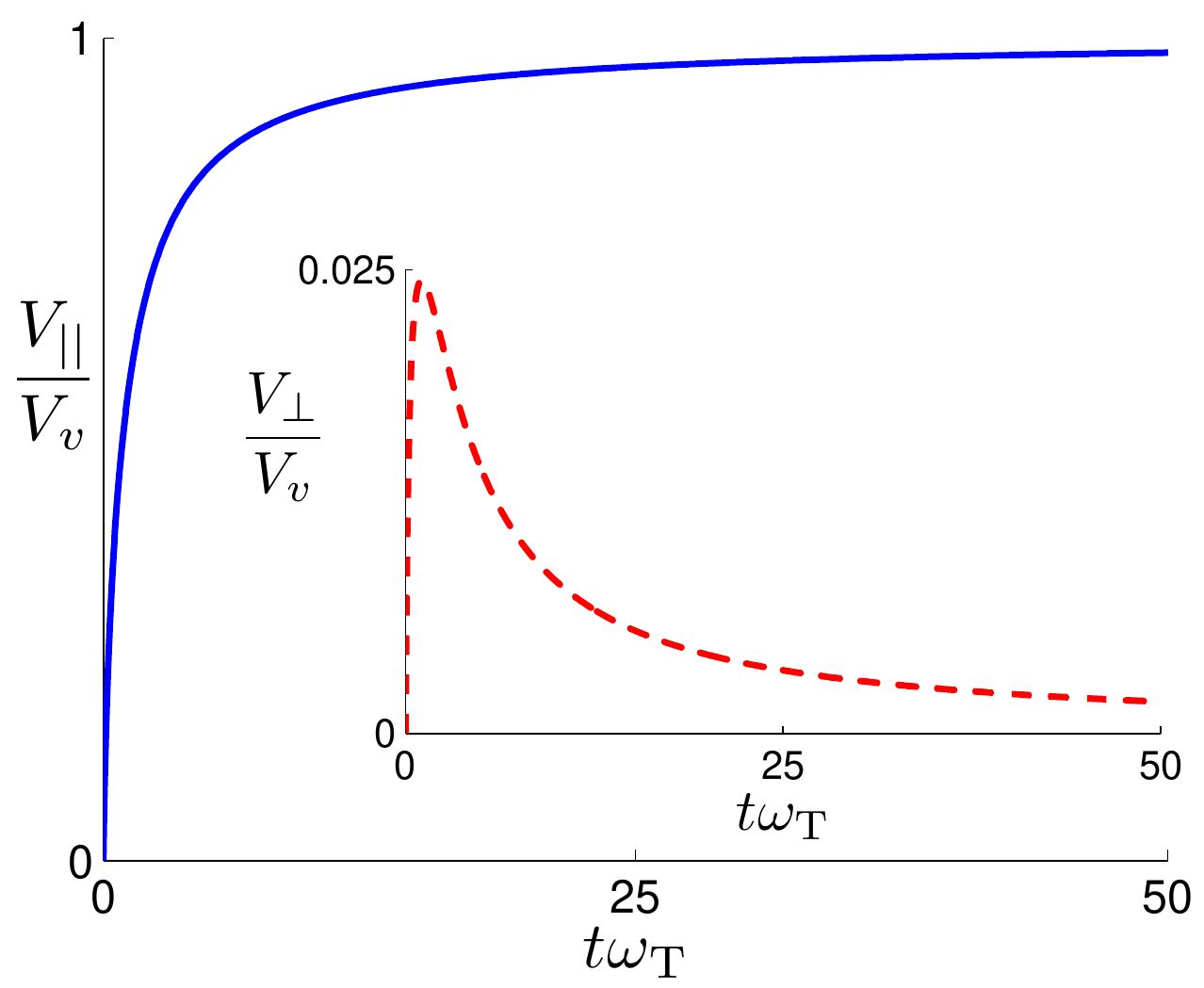}
\caption{The components of the vortex velocity which are parallel (main figure) and perpendicular (inset) to its final velocity its $\vc{V}_v$ as a function of time $t$ after it is released from rest into a background super flow. Memory effects are included but not inertia. Parameters are chosen so that  $\gamma_0/\rho\kappa=0.1$. \label{fig:vllvstmem}}

\end{figure}

Because the kernels $\gamma_{\parallel}(\omega)$ and $\gamma_R(\omega)$ are smooth functions of $\omega$ for all real frequencies except $\omega=0$, the long time behaviour of $\gamma(t)$ and functions like $\vc{G}(t)$ and $\vc{H}(t)$ (which are similarly smooth functions of $\omega$ and $\gamma_R(\omega)$) will be determined by the low-frequency behaviour of $\gamma_R(\omega)$. Since $\gamma_R(\omega)$ in this regime is a function of the dimensionless variable $\tilde{\omega} = \omega/\omega_T$, low frequencies here means $\tilde{\omega} \ll 1$. In this regime $\gamma_R(\omega)$ behaves as
\begin{widetext}
\begin{equation}
 \label{eq:lfrgamma}
\gamma_R(\tilde{\omega})\;\sim\; D_0(T)\left\{1\;+\;\frac{i\zeta(3)}{4\pi\zeta(4)} \omega
\left[\ln\left(\frac{\tilde{\omega}-i\delta}{\xi}\right)+
\ln\left(\frac{-\tilde{\omega}+i\delta}{\xi}\right)\right]\right\}\;+\;
\mathcal{O}\left(\tilde{\omega}^2
\ln\tilde{\omega}\right)\phantom{sp}\;\;(\textrm{for }\tilde{\omega}\ll 1)
\end{equation}
\end{widetext}
where the infinitesimal $\delta = 0^+$ ensures that the poles of $\gamma_R(\omega)$ lie in the upper half of the complex plane. $\xi$ is a dimensionless constant which is unimportant for determining the leading order long time behaviour of $\gamma_R(t)$ and related functions. Sometimes it will be convenient to write the expression \eqref{eq:lfrgamma} using the following abbreviation,
\begin{equation}
\gamma_R(\tilde\omega)\sim D_0(T)\left\{1-\frac{\zeta(3)}{2\zeta(4)}a_R(\tilde\omega)\right\} \;+\;\mathcal{O}
\left[\tilde{\omega}^2
\ln \tilde{\omega}\right]
\end{equation}
where $a_R(\tilde\omega)$ is the retarded time version of the absolute value function $|\tilde\omega|$. Then for $\tilde{\omega} \ll 1$, the low frequency components of the Green function have the  form,
\begin{equation}
\vc{G}(\omega)\; \sim\;-\frac{1}{\omega}\left\{\vc{h}_0+\vc{h}_a
\frac{a_R(\tilde\omega)}{\omega_T}+\mathcal{O}
(\tilde\omega)\right\}
 \label{eq:gmemnoMlowf}
\end{equation}
where the coefficients $\vc{h}_0$ and $\vc{h}_a$ are tensors given by
\begin{align}
\vc{h}_0&=\frac{\gamma_0\boldsymbol{\sigma}_0-i\rho\kappa\boldsymbol{\sigma}_2}{\gamma_0^2+\rho^2\kappa^2}
 \label{eq:ho}\\
\vc{h}_a&=\frac{\zeta(3)}{2\zeta(4)}\frac{\gamma_0}{(\rho^2\kappa^2+\gamma_0^2)^2}
\Bigl[(\gamma_0^2-\rho^2\kappa^2)\boldsymbol{\sigma}_0-2i\gamma_0\rho\kappa\boldsymbol{\sigma}_2\Bigr]. \label{eq:ha}
\end{align}
Adding the memory term renders the Green functions non-analytic at $\omega=0$; instead of decaying exponentially they decay algebraically. Thus, for the velocity Green function we have
\begin{equation}\label{eq:longtimetailh}
\vc{H}(t)\sim -\frac{\vc{h}_a}{\pi}\frac{1}{\omega_T t^2}+\mathcal{O}(t^4)\,\textrm{ as }t \rightarrow \infinity.
\end{equation}
so that at low temperatures one has
\begin{equation}
\vc{H}(t)\sim\tfrac{3\zeta(3)}{2}\frac{m_0\omega_T^3}{\omega_c^3\rho\kappa{t^2}}
\boldsymbol{\sigma}_0+i\frac{9\zeta(3)\zeta(4)}{2}\frac{\pi m_0^2\omega_T^7}{\omega_c^6\rho^2\kappa^2t^2}\boldsymbol{\sigma}_2.
\end{equation}
and for the position Green's function $\vc{G}(t)$ one has a much slower decay towards the steady state:
\begin{equation}\label{eq:longtimetailg}
\vc{G}(t)\sim \frac{\vc{h}_0}{2}\mathrm{sgn}(t)-\frac{2\vc{h}_a}{\omega_T t}+\mathcal{O}(t^3)\,\;\;(\text{as }t \rightarrow \infinity)
\end{equation}
The high frequency behaviour of the Green functions is also changed. One has $\gamma_R(\omega)\rightarrow\gamma_\infinity=D_0(T)/16$ as $\omega/\omega_T\rightarrow\infinity$, so that in this limit,
\begin{equation}
{\vc{G}}(\omega)\sim\frac{-i}{\omega}\left(\frac{{{\gamma}_\infinity}
\boldsymbol{\sigma}_0-i{{\rho\kappa}}\boldsymbol{\sigma}_2} {{{\rho^2\kappa^2}}+\gamma_\infinity^2}\right)
\end{equation}
The time dependence of the velocity Green function is shown in figs. \ref{fig:Hllmem} and \ref{fig:Hllmem}. Both Green functions are monotonic functions of time; without inertia there are no oscillations, and the response to sudden changes is now a slow algebraic decay. Figure \ref{fig:vllvstmem} illustrates the $\sim t^{-1}$ decay of the vortex velocity to its steady state velocity ${\bf V}_v$.

\begin{figure}
\includegraphics[width=8.6cm]{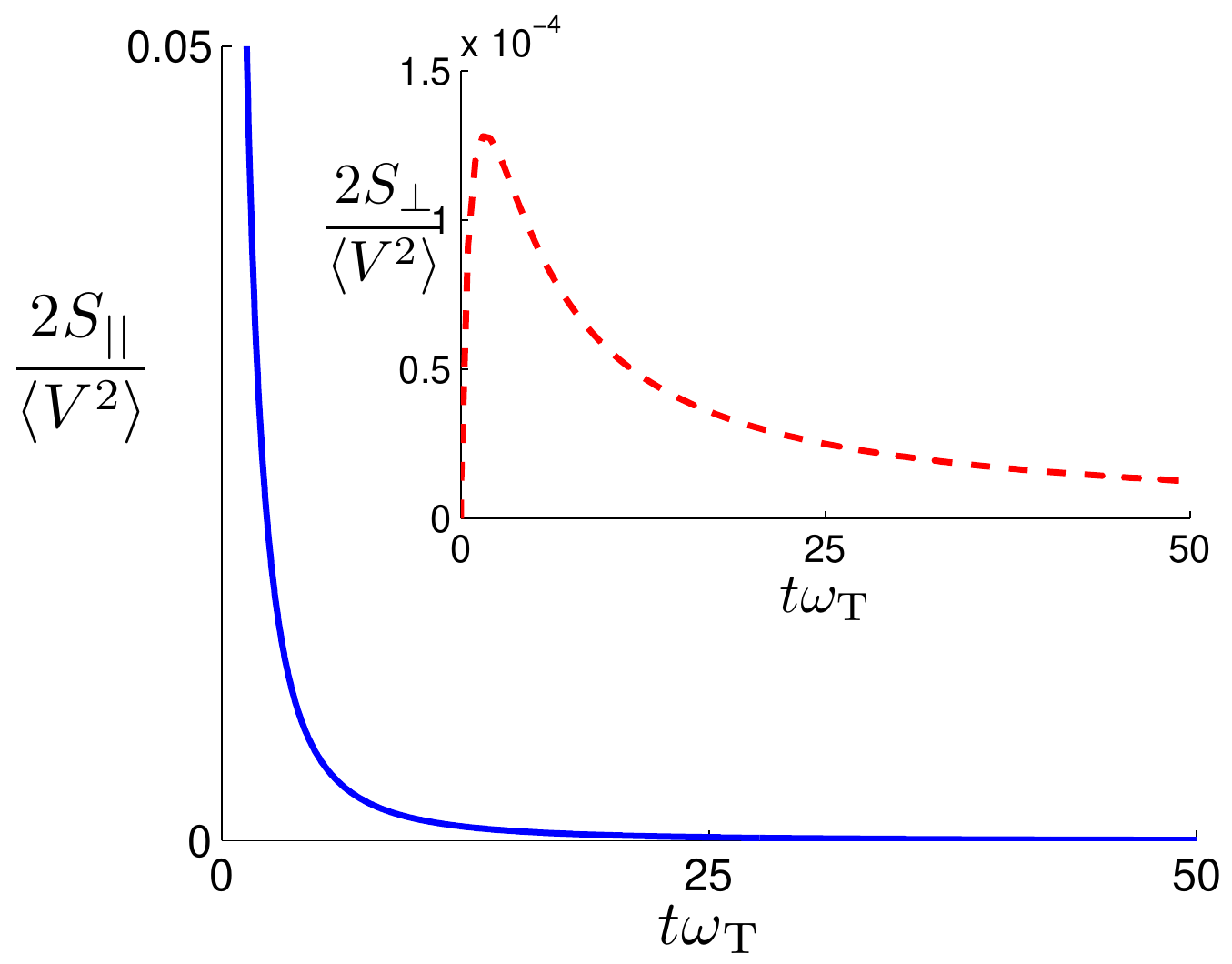}
\caption{The effect of the memory forces on the decay of velocity correlations in the inertialess case. The longitudinal (main figure) and transverse (inset) components of velocity-velocity correlator $\mathbf{S}_{\mathbf{vv}}$ are plotted. Parameters are chosen so that  $\gamma_0/\rho\kappa=0.1$.\label{fig:Sllmem}}

\end{figure}

Turning now to the full stochastic dynamics for this case, we note that the inclusion of memory effects in the friction kernel means, for consistency, that we must also include memory effects in the fluctuation noise source term. Thus, to evaluate the stochastic equations for the position and velocity correlators, we need the low frequency
($\tilde\omega\ll 1$) asymptotic forms of $\tilde{\chi}_\parallel(\omega)$ and $\tilde{\chi}_\perp(\omega)$, which are
\begin{equation}
\chi_\parallel(\tilde\omega) \;\sim\; \chi_0(T)\left(1-\frac{\zeta(3)}{2\zeta(4)}|\tilde\omega|\right)+
\mathcal{O}(\tilde\omega^2)
\end{equation}
for the longitudinal noise correlator, and
\begin{equation}
\chi_\perp(\tilde\omega) \;\sim\; \chi_0(T)\frac{\omega}{2\omega_\mathrm{c}}\left(1\left[1+\frac{\zeta(3)}{8\zeta(4)}\right]
|\tilde\omega|\right)+\mathcal{O}(\tilde\omega^3)
\end{equation}
for the transverse correlator. Again we see the non-analyticity of these forms in the $\tilde\omega \rightarrow 0$ limit. The velocity-velocity correlation function is also non-analytic at low frequency, where its asymptotic form is
\begin{widetext}
\begin{equation}
\vc{S}_\vc{vv}(\omega)\;\;\sim\;\; \vc{S}_\vc{vv}(0)-\frac{\zeta(3)\chi_0\rho\kappa}{2\zeta(4)(\rho^2\kappa^2+\gamma_0^2)^2
\omega_T}\left\{\rho\kappa|\omega|\boldsymbol{\sigma}_0+i\gamma_0(a_R(\omega)-
a_R(-\omega))\boldsymbol{\sigma}_2\right\}\,\;\;\;\;\;(\textrm{for }\omega\rightarrow 0).
\end{equation}
Therefore the  leading order algebraic tail of $\vc{S}_\vc{vv}(t)$ is
\begin{equation}\label{eq:svvltt}
\vc{S}_\vc{vv}(t)\;\;\sim\;\; \frac{\zeta(3)\chi_0\rho\kappa}{2\zeta(4)\pi(\rho^2\kappa^2+\gamma_0^2)^2
\omega_T}\left\{\frac{\rho\kappa}{t^2}\boldsymbol{\sigma}_0-
i\frac{\gamma_0\mathrm{sgn}(t)}{t^2}\boldsymbol{\sigma}_2
\right\}\,\qquad (\textrm{ for }t\rightarrow \infinity)
\end{equation}
\end{widetext}
where at low temperatures, the longitudinal and transverse components scale like $T^4$ and $T^8$ respectively. The high $\omega/\omega_T$ limits of both of the auto-correlators ${S}_\vc{vv}(\omega)$ and $\tilde{S}_\vc{RR}(\omega)$ are changed from the memoryless case with a replacements $\chi_0\rightarrow\chi_\infty=\zeta(5)\chi_0(T)/4\zeta(4)$ and $\gamma_0\rightarrow\gamma_\infty$. The time dependence of the velocity-velocity correlation function is depicted in figure \ref{fig:Sllmem} .

\section{Solution to full TS equations in the Quantum Regime}\label{sec:HVImemin}


We now give the solution to the full TS equations (again, in the approximation where we ignore the frequency dependence of the transverse drag term $\gamma_{\perp}(\omega, T)$). Thus we now include inertial effects, friction effects with memory, and fluctuaion noise effects with memory, in addition to the terms in the standard HVI equations.

When memory effects and inertia are included the full forms \eqref{eq:fullG} and \eqref{eq:fullH} for the Green functions need to be considered. In general the memory kernel will shift the poles at $\omega_\pm$ of the Green's functions, but so long as temperature is low enough this will be a small shift. The leading order long-time behaviour of both of Green functions will remain unchanged from the massless case just discussed (compare eqtns. (\ref{eq:longtimetailh})-(\ref{eq:longtimetailg})); including inertial mass adds no extra non-analytic term at low frequencies. At high frequencies ($\omega_c\gg\omega\gg\omega_T$) the leading order behaviour of both Green functions is the same as in the memoryless case (equation \eqref{eq:highfrg}). The high frequency auto-correlation functions are as in equations \eqref{eq:srrhif} and \eqref{eq:svvhif} but with the replacement $\chi_0\rightarrow\chi_\infty$. The mean squared vortex velocity can be calculated using the same device as in the memoryless case (compare eqtn. (\ref{Svv-0+})), except using the full form (\ref{eq:fullH}) for the velocity Green function.

We can get an approximate form for the velocity auto-correlation function using the following approximations for $\chi_\parallel(\omega)$ and $\gamma(\omega)$ (here we drop $\chi_\perp(\omega)$ because it contains an extra factor of the compressibility):
\begin{align}
\chi^{\mathrm{apx}}(\omega)&=\chi_0\left(\frac{1+X_1|\omega|}{1+X_2|\omega|}\right)\\
\gamma^{\mathrm{apx}}(\omega)&=\gamma_0\left(\frac{1+Y_1|\omega|}{1+Y_2|\omega|}\right)
\end{align}

$X_1$ and $X_2\,(Y_1$ and $Y_2)$ are chosen so that the large $\omega/\omega_T$ behaviour of $\chi^{\mathrm{apx}}(\omega)\, (\gamma^{\mathrm{apx}}(\omega))$ matches that of $\chi_\parallel(\omega),(\gamma(\omega))$; this means that we have the following approximate ratios:
\begin{align}
X_1&= \frac{\zeta(5)}{4\zeta(4)}X_2\approx\frac{0.1550}{\omega_T}  \\
X_2&= \frac{4\zeta(4)-\zeta(5)}{5\zeta(6)\omega_T}\approx\frac{0.6472}{\omega_T}
\end{align}
and
\begin{align}
Y_1&= \frac{5\zeta(4)}{64\zeta(5)\omega_T}\approx\frac{0.08155}{\omega_T}  \\
Y_2&=\frac{5\zeta(4)}{4\zeta(5)\omega_T}\approx\frac{1.305}{\omega_T}.
\end{align}
The retarded part of $\gamma^\mathrm{apx}(z)$ is then
\begin{equation}
\gamma_R^\mathrm{apx}(z)=\frac{Y_1}{2Y_2}+\frac{Y_2-Y_1}{\pi Y_2}\left[\frac{\ln(zY_2-i\epsilon)}{1 + (izY_2+\epsilon)^2}\right]
\end{equation}
where $\epsilon\rightarrow 0+$.

With these approximate forms we can then calculate an explicit result for the velocity auto-correlation function, written in the form
\begin{widetext}
\begin{align}
\left\langle\vc{v}(s+t)\otimes\vc{v}(s)\right\rangle&=\int_{-\infinity}^\infinity
\frac{\ud\omega}{2\pi}\chi^{\mathrm{apx}}(\omega)\vc{H}(\omega)\vc{H}^T(-\omega)e^{i\omega t}\\&=\chi_0\int_{0}^\infinity\frac{\ud\omega}{2\pi}
\left(\frac{1+X_1\omega}{1+X_2\omega}\right)\Bigl\{\vc{H}(\omega)\vc{H}^T(-\omega)e^{i\omega t}+\vc{H}^T(\omega)\vc{H}(-\omega)e^{-i\omega t}\Bigr\}.\label{eq:apxvvc}
\end{align}
Along the real line the integrand \eqref{eq:apxvvc} can be approximated by replacing $\vc{H}(\omega)$ and $\vc{H}^T(-\omega)$  by rational functions whose poles can be obtained by analytically continuing $\vc{H}(\omega)$ and $\vc{H}^T(-\omega)$  away from the real line. When $\vc{H}(\omega)$ is analytically continued away from the real line it has poles at $\omega = \omega_{\pm}$ where
\begin{equation}
\omega_\pm\;\;\;=\;\;\; \pm \Omega_o\left(1+ {\gamma_0 \over M_v} {(Y_2-Y_1)
\ln Y_2\Omega_o  \over Y_2^2\Omega_o^2- 1}\right)+i\frac{\gamma_0}{2M_v}
\frac{1+Y_1\Omega_o}{1+X_2\Omega_o}  \;\;\;\;\equiv\;\;\;\; \pm \tilde{\Omega}+i{{\gamma}^\mathrm{apx}(\Omega_o) \over 2M_v}.\label{eq:polesH}
\end{equation}
This expression is at next-to leading order in the compressibility so that the poles do not lie on the real axis. The ``cyclotron frequency'' of the pole is shifted from the memoryless result $\omega_o = \rho\kappa/M_v$ to $\tilde\Omega$, and the dissipation factor $\frac{{\gamma}^\mathrm{apx}(
\Omega_o)}{2M_v}$ is renormalised from the memoryless result $\tau_D^{-1}/2 = \gamma_0/2M_v$ by the frequency dependence of ${\gamma}(\omega)$. $\vc{H}(-\omega)$ has poles at $\omega=-\omega_\pm$ so that the velocity auto-correlator can be approximated by
\begin{align}
\left\langle\vc{v}(s+t)\otimes\vc{v}(s)\right\rangle \;\;=\;\;\chi_0\int_0^\infty\frac{\ud\omega}{2\pi}\left\{\vc{R}_{-X_2^{-1}}\frac{e^{i\omega t}}{\omega+X_2^{-1}}+\vc{R}^T_{-X_2^{-1}}\frac{e^{-i\omega t}}{\omega+X_2^{-1}}+\sum_{\omega_p}\vc{R}_{p}\frac{e^{i\omega t}}{\omega-\omega_p}+\sum_{\omega_p}\vc{R}^{\dagger}_{p}\;\frac{e^{-i\omega t}}{\omega-\omega_p}\right\}
 \label{vvR-pm}
\end{align}
where the sums contain terms from all the poles of $\vc{H}(\omega)\vc{H}^T(-\omega)$, $\omega_p$ runs over the 4 values $\omega_p = \omega_{\pm},-\omega_{\pm}$, and $\vc{R}_p$ is the residue for these 4 values, viz.,
\begin{equation}
\vc{R}_p \;\;=\;\;\mathrm{Res}\left(\frac{1+X_1\omega}{1+X_2\omega}\vc{H}(\omega)
\vc{H}^T(-\omega)\right)_{\omega = \omega_p}.
\end{equation}
We have also written $\vc{R}_p^{\dagger}$ to denote the conjugate transpose of $\vc{R}_p$. We can write $\vc{R}_p$ in more explicit form as
\begin{align}
\vc{R}_{p_\pm}\;\;=\;\; R_p^\parallel\boldsymbol{\sigma}_0-iR_p^\perp\boldsymbol{\sigma}_2 \;\;&=\;\;
\frac{1}{M^4}\left(\frac{1+X_1\omega_\pm}{1+X_2\omega_\pm}\right)\left[\frac{(\omega_\pm^2M^2+\Omega^2)
\boldsymbol{\sigma}_0-2\omega_\pm M\Omega\boldsymbol{\sigma}_2}{2\omega_\pm(\omega_\pm-\omega_\mp)(\omega_\pm+\omega_\mp)}\right]\\
\vc{R}_{-\omega_\pm}\;\;&=\;\; \vc{R}_{p_\mp}^\dagger\\
\vc{R}_{-X_2^{-1}} \;\;&=\;\;\left(\frac{1-X_1X_2^{-1}}{M^4X_2}\right)
\left[\frac{(X_2^{-2}M^2+\Omega^2)\boldsymbol{\sigma}_0+2X_2^{-1}
M\Omega\boldsymbol{\sigma}_2}{(\omega_\pm^2-X_2^{-2})(\omega_\mp^2-X_2^{-2})}\right]
\end{align}
where $\dagger$ again denotes the conjugate transpose. Thus we get the following approximate expression for the velocity auto-correlator:
\begin{align}
\left\langle\vc{v}(s+t)\otimes\vc{v}(s)\right\rangle \;\;=\;\;\frac{\chi_0}{\pi}\real\biggl\{\vc{R}_{-X_2^{-1}}&
\left[g\left(\frac{t}{X_2}\right)+if\left(\frac{t}{X_2}\right)\right]+
\left[2g(\omega_+t){R}^{\parallel}_{-\omega_+}+2g(\omega_+^*t){R}^{\parallel}_{-\omega_-}\right]
\boldsymbol{\sigma}_0\biggr.\nonumber\\
&\biggl.+\left[2f(\omega_+t)R_{-\omega+}^\perp-2f(\omega_+^*t){R}_{-\omega_-}^\perp\right]\boldsymbol{\sigma}_2-2\pi i\vc{R}_{-\omega_-}^{\dagger}e^{i\omega_-t}\biggr\}\label{eq:vautohalfsimplified}
\end{align}
\end{widetext}
where $f(z)$ and $g(z)$ have the integral representations \cite{OLBC10},
\begin{align}
f(z)=\int_0^\infty\ud\xi\frac{\sin\xi}{\xi+z} \; ; \qquad
g(z)=\int_0^\infty\ud\xi\frac{\cos\xi}{\xi+z}.
\end{align}
ie., they are the auxiliary functions for the sine and cosine integrals.

The results in (\ref{eq:vautohalfsimplified}) and (\ref{vvR-pm}), based on (\ref{eq:apxvvc}), are still quite formal - what we wish to know is the qualitative behaviour. Notice first that the last term in the expression \eqref{eq:vautohalfsimplified} for the velocity auto-correlator is dominant at low compressibility and for times $t\ll \frac{M_v}{\gamma_0}$, in which case
\begin{widetext}
\begin{equation}
\left\langle\vc{v}(s+t)\otimes\vc{v}(s)\right\rangle \;\;\sim \;\;\frac{{\chi}^\mathrm{apx}\left(\frac{\Omega}{M_v}\right)}{2M_v}
\frac{\exp\left[-\frac{{\gamma}^{\mathrm{apx}}\left(\frac{\Omega}{M_v}\right)t}{M_v}\right]}
{{\gamma}^{\mathrm{apx}}\left(\frac{\Omega}{M_v}\right)}
\left[\cos\left(\frac{\tilde{\Omega}t}{M_v}\right)\boldsymbol{\sigma}_0+
i\sin\left(\frac{\tilde{\Omega}t}{M_v}\right)\boldsymbol{\sigma}_2\right].
 \label{eq:locompvv}
\end{equation}
\end{widetext}
Algebraic decay tails do not occur at leading order in the compressibility, and so the leading order equilibrium mean squared velocity is (compare to equation \eqref{eq:msv}):
\begin{equation}
\left\langle\vc{v}(t)\otimes\vc{v}(t)\right\rangle \;\;\sim \;\;\frac{{\chi}^\mathrm{apx}
\left(\frac{\Omega}{M_v}\right)}{2M_v{\gamma}^{\mathrm{apx}}
\left(\frac{\Omega}{M_v}\right)}\boldsymbol{\sigma}_0.
 \label{eq:avv^2withmem}
\end{equation}
We can see that the next to leading order terms in compressibility contain a contribution to the algebraic decay tail of the velocity auto-correlator, because for large $t$ and arg$z<\pi$ the functions $g(zt)$ and $f(zt)$ have the following asymptotic behaviour:
\begin{align}
g(zt)& \;\sim \;\frac{1}{zt}\left[1-\frac{2!}{z^2t^2}+\frac{4!}{z^4t^4}+\mathcal{O}
\left(z^{-6}t^{-6}\right)\right]\\
f(zt)& \;\sim \; \frac{1}{z^2t^2}\left[1-\frac{3!}{z^2t^2}+\frac{5!}{z^4t^4}+
\mathcal{O}\left(z^{-6}t^{-6}\right)\right].
\end{align}
After enough time has passed by, the contribution from the algebraic decay tail \eqref{eq:svvltt} will always dominate the lowest order in compressibility solution in \eqref{eq:locompvv}; the crossover between the two terms will happen when the time is of order $t_c$, with $t_c$ given by
\begin{align}
t_c \;=\;\frac{-M_v}{{\gamma}^\mathrm{apx}\left(\frac{\Omega}{M_v}\right)}\mathrm{Wm}
\left(-\sqrt{\frac{\zeta(3)}{4\pi}\frac{\chi_0\left[{\gamma}^\mathrm{apx}
\left(\frac{\Omega}{M_v}\right)\right]^3}{ \chi^\mathrm{apx}\left(\frac{\Omega}{M_v}\right)\omega_T M_v}}\right)
\end{align}
where $\mathrm{Wm}$ is that branch of the Lambert W-function which maps the domain $\left[-\frac{1}{e},0\right)$ to the range $(-\infty,-1]$. For small enough temperatures $t_c$ can then be approximated by
\begin{equation}
t_c \;\;\sim\;\; \frac{M_v}{2\tilde{\gamma}^\mathrm{apx}(\frac{\Omega}{M_v})}
\left\{\log\left(\frac{\chi^\mathrm{apx}\left(\frac{\Omega}{M_v}\right)\omega_T M_v}{\chi_0\left[{\gamma}^\mathrm{apx}
\left(\frac{\Omega}{M_v}\right)\right]^3}\right)-3.119\right\}
\end{equation}
Thus the crossover time is given by the dissipation time scale weighted by terms that slowly vary with the ratios between the dissipation time scale and the cyclotron orbit period, the cyclotron frequency and the thermal frequency, and the bare and renormalised noise strength. When viewed on time scales larger than $t_c$ the underdamped cyclotron motion of the vortex is rapidly dissipated by friction so all that is visible is the slow decay from that present in the inertia free case.

The steady state average vortex velocity in the presence of a steady state background flow can also be calculated, via a similar calculation to that above; one finds that the velocity tends to that given in expression \eqref{eq:avvHVI}, but with the friction coefficient renormalised to the real part of its value at the pole, ie., $\gamma_0\rightarrow {\gamma}^\mathrm{apx}(\Omega/M_v)$.

\section{Conclusions}

We have considered the effect of memory forces and inertia on the the motion of a single quantum vortex in a homogeneous Bose superfluid, using the TS quantum equations of motion for the vortex. Because of the vortex inertia, the otherwise instantaneous reaction of a vortex to applied forces is replaced by damped oscillations at a renormalized cyclotron frequency. The simultaneous presence of noise and inertia leads over long time periods to diffusive vortex dynamics. The inclusion of memory effects in the quasiparticle forces on the vortex, as well as memory effects in the noise correlator changes this again: the response to an applied force then has a component that decays algebraically rather than exponentially, and the parameters determining the diffusive behaviour are renormalised compared to the memoryless case.

Quantum vortices occur in a large range of superfluids including liquid $^4$He \cite{Donnelly91}, Bose condensed gases \cite{FetterSvidzinsky}, and superconductors \cite{Blatter94}. 
In any application to real superfluids, the validity of the TS equations of motion, and of our results, requires that both the characteristic thermal frequency $\omega_T$ and the frequency of vortex motion be much less the the characteristic compressional frequency $\omega_\mathrm{c}$. It also requires that we deal with length scales much greater than the vortex core radius. 

These conditions prevail for slowly moving vortices in liquid $^4$He at temperatures  $T\lesssim 0.7K$. One problem here is that individual vortices in superfluid $^4$He are hard to observe - one requires some sort of tracer particle which is attracted to the vortex core \cite{Bewley06}, \cite{Yarmchuk79}. The problem here is that the particle can change the vortex dynamics significantly (in the case of an electron tracer, the change is drastic). 

In Bose condensed gases the situation is probably more favourable; anisotropic trapping potentials can be used to create quasi-two dimensional condensates containing quantum vortices \cite{Stock05,Hadzibabic06}, and individual vortices have been tracked dynamically and viewed in situ \cite{Freilich10,Wilson14}. Moreover, the properties of the quantum fluid itself may be modified over a large parameter range in a trapped gas, for instance by varying the strength of interactions via the Feshbach resonance \cite{Bloch08,Wilson14}. One complication comes from the inhomogeneity of the gas; the density gradient  creates an additional force on the vortex \cite{Svidzinsky00}, as well as changing the flow inside the droplet and causing an outward Magnus force \cite{FedichevShlyapnikov}. 

In spite of these complications, we expect the main results given here to apply, although it is clear that one requires specific calculations for well-defined geometries if one is to make any experimental predictions. Thus, for example, we expect the vortex velocity will still have an algebraic decay due to the non-analytic memory kernel, giving a dynamics quite different from that predicted using the HVI equations. This is quite apart from differences caused by vortex inertia.

\section{Acknowledgements}

We thank Julien Froustey and Lara Thompson in Vancouver, and Jean Dalibard and his group in Paris, for discussions. This work was supported by an NSERC discovery grant in Canada, and also benefited from a an extended visit to Paris by PCES, as part of a College de France professorship there.

\providecommand{\noopsort}[1]{}\providecommand{\singleletter}[1]{#1}%


\begin{thebibliography}{31}
\expandafter\ifx\csname natexlab\endcsname\relax\def\natexlab#1{#1}\fi
\expandafter\ifx\csname bibnamefont\endcsname\relax
  \def\bibnamefont#1{#1}\fi
\expandafter\ifx\csname bibfnamefont\endcsname\relax
  \def\bibfnamefont#1{#1}\fi
\expandafter\ifx\csname citenamefont\endcsname\relax
  \def\citenamefont#1{#1}\fi
\expandafter\ifx\csname url\endcsname\relax
  \def\url#1{\texttt{#1}}\fi
\expandafter\ifx\csname urlprefix\endcsname\relax\def\urlprefix{URL }\fi
\providecommand{\bibinfo}[2]{#2}
\providecommand{\eprint}[2][]{\url{#2}}

\bibitem[{\citenamefont{Hall and Vinen}(1956)}]{HallVinen}
\bibinfo{author}{\bibfnamefont{H.~E.} \bibnamefont{Hall}} \bibnamefont{and}
  \bibinfo{author}{\bibfnamefont{W.~F.} \bibnamefont{Vinen}},
  \bibinfo{journal}{Proceedings of the Royal Society of London A: Mathematical,
  Physical and Engineering Sciences} \textbf{\bibinfo{volume}{238}},
  \bibinfo{pages}{204} (\bibinfo{year}{1956}), ISSN \bibinfo{issn}{0080-4630},

\bibitem[{\citenamefont{Iordansky}(1964)}]{Iordanskii64}
\bibinfo{author}{\bibfnamefont{S.}~\bibnamefont{Iordansky}},
  \bibinfo{journal}{Annals of Physics} \textbf{\bibinfo{volume}{29}},
  \bibinfo{pages}{335} (\bibinfo{year}{1964}).

\bibitem[{\citenamefont{Sonin}(1987)}]{Sonin87}
\bibinfo{author}{\bibfnamefont{E.~B.} \bibnamefont{Sonin}},
  \bibinfo{journal}{Rev. Mod. Phys.} \textbf{\bibinfo{volume}{59}},
  \bibinfo{pages}{87} (\bibinfo{year}{1987}),

\bibitem[{\citenamefont{Blatter et~al.}(1994)\citenamefont{Blatter, Feigel'man,
  Geshkenbein, Larkin, and Vinokur}}]{Blatter94}
\bibinfo{author}{\bibfnamefont{G.}~\bibnamefont{Blatter}},
  \bibinfo{author}{\bibfnamefont{M.~V.} \bibnamefont{Feigel'man}},
  \bibinfo{author}{\bibfnamefont{V.~B.} \bibnamefont{Geshkenbein}},
  \bibinfo{author}{\bibfnamefont{A.~I.} \bibnamefont{Larkin}},
  \bibnamefont{and} \bibinfo{author}{\bibfnamefont{V.~M.}
  \bibnamefont{Vinokur}}, \bibinfo{journal}{Rev. Mod. Phys.}
  \textbf{\bibinfo{volume}{66}}, \bibinfo{pages}{1125} (\bibinfo{year}{1994}),

\bibitem[{\citenamefont{Sonin}(1997)}]{Sonin97}
\bibinfo{author}{\bibfnamefont{E.~B.} \bibnamefont{Sonin}},
  \bibinfo{journal}{Phys. Rev. B} \textbf{\bibinfo{volume}{55}},
  \bibinfo{pages}{485} (\bibinfo{year}{1997}),

\bibitem[{\citenamefont{Kopnin}(2002)}]{Kopnin02}
\bibinfo{author}{\bibfnamefont{N.}~\bibnamefont{Kopnin}},
  \bibinfo{journal}{Reports on Progress in Physics}
  \textbf{\bibinfo{volume}{65}}, \bibinfo{pages}{1633} (\bibinfo{year}{2002}).

\bibitem[{\citenamefont{Thouless et~al.}(1996)\citenamefont{Thouless, Ao, and
  Niu}}]{ThoulessI}
\bibinfo{author}{\bibfnamefont{D.~J.} \bibnamefont{Thouless}},
  \bibinfo{author}{\bibfnamefont{P.}~\bibnamefont{Ao}}, \bibnamefont{and}
  \bibinfo{author}{\bibfnamefont{Q.}~\bibnamefont{Niu}},
  \bibinfo{journal}{Phys. Rev. Lett.} \textbf{\bibinfo{volume}{76}},
  \bibinfo{pages}{3758} (\bibinfo{year}{1996}),

\bibitem[{\citenamefont{Thouless and Anglin}(2007)}]{ThoulessII}
\bibinfo{author}{\bibfnamefont{D.~J.} \bibnamefont{Thouless}} \bibnamefont{and}
  \bibinfo{author}{\bibfnamefont{J.~R.} \bibnamefont{Anglin}},
  \bibinfo{journal}{Phys. Rev. Lett.} \textbf{\bibinfo{volume}{99}},
  \bibinfo{pages}{105301} (\bibinfo{year}{2007}),

\bibitem[{\citenamefont{Popov}(1973)}]{Popov73}
\bibinfo{author}{\bibfnamefont{V.}~\bibnamefont{Popov}},
  \bibinfo{journal}{Soviet Journal of Experimental and Theoretical Physics}
  \textbf{\bibinfo{volume}{37}}, \bibinfo{pages}{341} (\bibinfo{year}{1973}).

\bibitem[{\citenamefont{Duan}(1994)}]{Duan94}
\bibinfo{author}{\bibfnamefont{J.-M.} \bibnamefont{Duan}},
  \bibinfo{journal}{Phys. Rev. B} \textbf{\bibinfo{volume}{49}},
  \bibinfo{pages}{12381} (\bibinfo{year}{1994}),

\bibitem[{\citenamefont{Arovas and Freire}(1997)}]{Arovas97}
\bibinfo{author}{\bibfnamefont{D.~P.} \bibnamefont{Arovas}} \bibnamefont{and}
  \bibinfo{author}{\bibfnamefont{J.}~\bibnamefont{Freire}},
  \bibinfo{journal}{Phys. Rev. B} \textbf{\bibinfo{volume}{55}},
  \bibinfo{pages}{1068} (\bibinfo{year}{1997}),

\bibitem[{\citenamefont{Wang et~al.}(2010)\citenamefont{Wang, Duine, and
  MacDonald}}]{MacDonald10}
\bibinfo{author}{\bibfnamefont{C.-C.~J.} \bibnamefont{Wang}},
  \bibinfo{author}{\bibfnamefont{R.~A.} \bibnamefont{Duine}}, \bibnamefont{and}
  \bibinfo{author}{\bibfnamefont{A.~H.} \bibnamefont{MacDonald}},
  \bibinfo{journal}{Phys. Rev. A} \textbf{\bibinfo{volume}{81}},
  \bibinfo{pages}{013609} (\bibinfo{year}{2010}),

\bibitem[{\citenamefont{Volovik}(1996)}]{Volovik}
\bibinfo{author}{\bibfnamefont{G.~E.} \bibnamefont{Volovik}},
  \bibinfo{journal}{Phys. Rev. Lett.} \textbf{\bibinfo{volume}{77}},
  \bibinfo{pages}{4687} (\bibinfo{year}{1996}),

\bibitem[{\citenamefont{Sonin}(1998)}]{Sonin98}
\bibinfo{author}{\bibfnamefont{E.~B.} \bibnamefont{Sonin}},
  \bibinfo{journal}{Phys. Rev. Lett.} \textbf{\bibinfo{volume}{81}},
  \bibinfo{pages}{4276} (\bibinfo{year}{1998}),

\bibitem[{\citenamefont{Wexler et~al.}(1998)\citenamefont{Wexler, Thouless, Ao,
  and Niu}}]{Wexler98}
\bibinfo{author}{\bibfnamefont{C.}~\bibnamefont{Wexler}},
  \bibinfo{author}{\bibfnamefont{D.~J.} \bibnamefont{Thouless}},
  \bibinfo{author}{\bibfnamefont{P.}~\bibnamefont{Ao}}, \bibnamefont{and}
  \bibinfo{author}{\bibfnamefont{Q.}~\bibnamefont{Niu}},
  \bibinfo{journal}{Phys. Rev. Lett.} \textbf{\bibinfo{volume}{81}},
  \bibinfo{pages}{4277} (\bibinfo{year}{1998}),

\bibitem[{\citenamefont{Thompson and Stamp}(2012)}]{TS1}
\bibinfo{author}{\bibfnamefont{L.}~\bibnamefont{Thompson}} \bibnamefont{and}
  \bibinfo{author}{\bibfnamefont{P.~C.~E.} \bibnamefont{Stamp}},
  \bibinfo{journal}{Phys. Rev. Lett.} \textbf{\bibinfo{volume}{108}},
  \bibinfo{pages}{184501} (\bibinfo{year}{2012}),

\bibitem[{\citenamefont{Thompson and Stamp}(2013)}]{TS2}
\bibinfo{author}{\bibfnamefont{L.}~\bibnamefont{Thompson}} \bibnamefont{and}
  \bibinfo{author}{\bibfnamefont{P.~C.~E.} \bibnamefont{Stamp}},
  \bibinfo{journal}{Journal of Low Temperature Physics}
  \textbf{\bibinfo{volume}{171}}, \bibinfo{pages}{526} (\bibinfo{year}{2013}),
  ISSN \bibinfo{issn}{1573-7357},

\bibitem[{\citenamefont{Haldane and Wu}(1985)}]{Haldane85}
\bibinfo{author}{\bibfnamefont{F.~D.~M.} \bibnamefont{Haldane}}
  \bibnamefont{and} \bibinfo{author}{\bibfnamefont{Y.-S.} \bibnamefont{Wu}},
  \bibinfo{journal}{Phys. Rev. Lett.} \textbf{\bibinfo{volume}{55}},
  \bibinfo{pages}{2887} (\bibinfo{year}{1985}),

\bibitem[{\citenamefont{Ao and Thouless}(1993)}]{Ao93}
\bibinfo{author}{\bibfnamefont{P.}~\bibnamefont{Ao}} \bibnamefont{and}
  \bibinfo{author}{\bibfnamefont{D.~J.} \bibnamefont{Thouless}},
  \bibinfo{journal}{Phys. Rev. Lett.} \textbf{\bibinfo{volume}{70}},
  \bibinfo{pages}{2158} (\bibinfo{year}{1993}),

\bibitem[{\citenamefont{Donnelly}(1991)}]{Donnelly91}
\bibinfo{author}{\bibfnamefont{R.~J.} \bibnamefont{Donnelly}},
  \emph{\bibinfo{title}{Quantized vortices in helium II}},
  vol.~\bibinfo{volume}{2} (\bibinfo{publisher}{Cambridge University Press},
  \bibinfo{year}{1991}).

\bibitem[{\citenamefont{Olver}(2010)}]{OLBC10}
\bibinfo{author}{\bibfnamefont{F.~W.} \bibnamefont{Olver}},
  \emph{\bibinfo{title}{NIST Handbook of Mathematical Functions Hardback and
  CD-ROM}} (\bibinfo{publisher}{Cambridge University Press},
  \bibinfo{year}{2010}).

\bibitem[{\citenamefont{Fetter and Svidzinsky}(2001)}]{FetterSvidzinsky}
\bibinfo{author}{\bibfnamefont{A.~L.} \bibnamefont{Fetter}} \bibnamefont{and}
  \bibinfo{author}{\bibfnamefont{A.~A.} \bibnamefont{Svidzinsky}},
  \bibinfo{journal}{Journal of Physics: Condensed Matter}
  \textbf{\bibinfo{volume}{13}}, \bibinfo{pages}{R135} (\bibinfo{year}{2001}),

\bibitem[{\citenamefont{Bewley et~al.}(2006)\citenamefont{Bewley, Lathrop, and
  Sreenivasan}}]{Bewley06}
\bibinfo{author}{\bibfnamefont{G.~P.} \bibnamefont{Bewley}},
  \bibinfo{author}{\bibfnamefont{D.~P.} \bibnamefont{Lathrop}},
  \bibnamefont{and} \bibinfo{author}{\bibfnamefont{K.~R.}
  \bibnamefont{Sreenivasan}}, \bibinfo{journal}{Nature}
  \textbf{\bibinfo{volume}{441}}, \bibinfo{pages}{588} (\bibinfo{year}{2006}).

\bibitem[{\citenamefont{Yarmchuk et~al.}(1979)\citenamefont{Yarmchuk, Gordon,
  and Packard}}]{Yarmchuk79}
\bibinfo{author}{\bibfnamefont{E.~J.} \bibnamefont{Yarmchuk}},
  \bibinfo{author}{\bibfnamefont{M.~J.~V.} \bibnamefont{Gordon}},
  \bibnamefont{and} \bibinfo{author}{\bibfnamefont{R.~E.}
  \bibnamefont{Packard}}, \bibinfo{journal}{Phys. Rev. Lett.}
  \textbf{\bibinfo{volume}{43}}, \bibinfo{pages}{214} (\bibinfo{year}{1979}),

\bibitem[{\citenamefont{Stock et~al.}(2005)\citenamefont{Stock, Hadzibabic,
  Battelier, Cheneau, and Dalibard}}]{Stock05}
\bibinfo{author}{\bibfnamefont{S.}~\bibnamefont{Stock}},
  \bibinfo{author}{\bibfnamefont{Z.}~\bibnamefont{Hadzibabic}},
  \bibinfo{author}{\bibfnamefont{B.}~\bibnamefont{Battelier}},
  \bibinfo{author}{\bibfnamefont{M.}~\bibnamefont{Cheneau}}, \bibnamefont{and}
  \bibinfo{author}{\bibfnamefont{J.}~\bibnamefont{Dalibard}},
  \bibinfo{journal}{Phys. Rev. Lett.} \textbf{\bibinfo{volume}{95}},
  \bibinfo{pages}{190403} (\bibinfo{year}{2005}),

\bibitem[{\citenamefont{Hadzibabic et~al.}(2006)\citenamefont{Hadzibabic,
  Kr{\"u}ger, Cheneau, Battelier, and Dalibard}}]{Hadzibabic06}
\bibinfo{author}{\bibfnamefont{Z.}~\bibnamefont{Hadzibabic}},
  \bibinfo{author}{\bibfnamefont{P.}~\bibnamefont{Kr{\"u}ger}},
  \bibinfo{author}{\bibfnamefont{M.}~\bibnamefont{Cheneau}},
  \bibinfo{author}{\bibfnamefont{B.}~\bibnamefont{Battelier}},
  \bibnamefont{and} \bibinfo{author}{\bibfnamefont{J.}~\bibnamefont{Dalibard}},
  \bibinfo{journal}{Nature} \textbf{\bibinfo{volume}{441}},
  \bibinfo{pages}{1118} (\bibinfo{year}{2006}).

\bibitem[{\citenamefont{Freilich et~al.}(2010)\citenamefont{Freilich, Bianchi,
  Kaufman, Langin, and Hall}}]{Freilich10}
\bibinfo{author}{\bibfnamefont{D.}~\bibnamefont{Freilich}},
  \bibinfo{author}{\bibfnamefont{D.}~\bibnamefont{Bianchi}},
  \bibinfo{author}{\bibfnamefont{A.}~\bibnamefont{Kaufman}},
  \bibinfo{author}{\bibfnamefont{T.}~\bibnamefont{Langin}}, \bibnamefont{and}
  \bibinfo{author}{\bibfnamefont{D.}~\bibnamefont{Hall}},
  \bibinfo{journal}{Science} \textbf{\bibinfo{volume}{329}},
  \bibinfo{pages}{1182} (\bibinfo{year}{2010}).

\bibitem[{\citenamefont{Wilson et~al.}(2015)\citenamefont{Wilson, Newman,
  Lowney, and Anderson}}]{Wilson14}
\bibinfo{author}{\bibfnamefont{K.~E.} \bibnamefont{Wilson}},
  \bibinfo{author}{\bibfnamefont{Z.~L.} \bibnamefont{Newman}},
  \bibinfo{author}{\bibfnamefont{J.~D.} \bibnamefont{Lowney}},
  \bibnamefont{and} \bibinfo{author}{\bibfnamefont{B.~P.}
  \bibnamefont{Anderson}}, \bibinfo{journal}{Phys. Rev. A}
  \textbf{\bibinfo{volume}{91}}, \bibinfo{pages}{023621}
  (\bibinfo{year}{2015}),

\bibitem[{\citenamefont{Bloch et~al.}(2008)\citenamefont{Bloch, Dalibard, and
  Zwerger}}]{Bloch08}
\bibinfo{author}{\bibfnamefont{I.}~\bibnamefont{Bloch}},
  \bibinfo{author}{\bibfnamefont{J.}~\bibnamefont{Dalibard}}, \bibnamefont{and}
  \bibinfo{author}{\bibfnamefont{W.}~\bibnamefont{Zwerger}},
  \bibinfo{journal}{Rev. Mod. Phys.} \textbf{\bibinfo{volume}{80}},
  \bibinfo{pages}{885} (\bibinfo{year}{2008}),

\bibitem[{\citenamefont{Svidzinsky and Fetter}(2000)}]{Svidzinsky00}
\bibinfo{author}{\bibfnamefont{A.~A.} \bibnamefont{Svidzinsky}}
  \bibnamefont{and} \bibinfo{author}{\bibfnamefont{A.~L.}
  \bibnamefont{Fetter}}, \bibinfo{journal}{Phys. Rev. A}
  \textbf{\bibinfo{volume}{62}}, \bibinfo{pages}{063617}
  (\bibinfo{year}{2000}),

\bibitem[{\citenamefont{Fedichev and Shlyapnikov}(1999)}]{FedichevShlyapnikov}
\bibinfo{author}{\bibfnamefont{P.~O.} \bibnamefont{Fedichev}} \bibnamefont{and}
  \bibinfo{author}{\bibfnamefont{G.~V.} \bibnamefont{Shlyapnikov}},
  \bibinfo{journal}{Phys. Rev. A} \textbf{\bibinfo{volume}{60}},
  \bibinfo{pages}{R1779} (\bibinfo{year}{1999}),

\end{thebibliography}
\end{document}